\documentclass{aa}  

\usepackage{graphicx}
\usepackage{txfonts}
\usepackage[colorlinks=true,
			linkcolor=black,
			citecolor=black,
			urlcolor=blue,
			]{hyperref}

\usepackage{ulem}

\newcommand{\RM}[1]{\MakeUppercase{\romannumeral #1{}}}
\begin{document} 

   \title{Young, metal-enriched cores in early-type dwarf galaxies in the Virgo cluster based on colour gradients}
   \titlerunning{Young, metal-enriched cores in early-type dwarf galaxies}
   %\subtitle{}

   \author{Linda Urich\inst{1}\fnmsep\thanks{\email{linda@dwarfgalaxies.net}},
          Thorsten Lisker\inst{1},
          Joachim Janz\inst{2,}\inst{3},
          Glenn van de Ven\inst{4},
          Ryan Leaman\inst{4},
          Alessandro Boselli\inst{5},
          Sanjaya Paudel\inst{6},
          Agnieszka Sybilska\inst{7},
          Reynier F. Peletier\inst{8},
          Mark den Brok\inst{9},
          Gerhard Hensler\inst{10},
          Elisa Toloba\inst{11}, 
          Jes\'us Falc\'on-Barroso\inst{12,}\inst{13}, 
          Sami-Matias Niemi\inst{14}
          }
   \authorrunning{L.\ Urich et al.}
   \institute{Astronomisches Rechen-Institut, Zentrum f\"ur Astronomie der Universit\"at Heidelberg, M\"onchhofstra\ss e 12-14, 69120 Heidelberg, Germany
            \and
            Centre for Astrophysics and Supercomputing, Swinburne University, Hawthorn, VIC 3122, Australia
            \and
            Division of Astronomy, Department of Physics, University of Oulu, P.O. Box 3000, FI-90014 Oulun Yliopisto, Finland 
            \and
            Max-Planck-Institut f\"ur Astronomie, K\"onigstuhl 17, 69117 Heidelberg, Germany
            \and
            Aix-Marseille Université, CNRS, LAM, Laboratoire d’Astrophysique de Marseille, 13388 Marseille, France 
            \and 
            Department of Astronomy \& Center for Galaxy Evolution Research, Yonsei University, Seoul 03722, Korea
            \and
            European Southern Observatory, Karl-Schwarzschild-Strasse 2, 85748 Garching bei M\"unchen, Germany
            \and
            Kapteyn Astronomical Institute, University of Groningen, Postbus 800, 9700 AV Groningen, the Netherlands
            \and
            Institute for Astronomy, ETH Zurich, Wolfgang-Pauli-Strasse 27, 8093 Zurich, Switzerland
            \and
            Department of Astrophysics, University of Vienna, Tuerkenschanzstr. 17, 1180 Vienna, Austria
            \and
            University of the Pacific, Department of Physics, 3601 Pacific Avenue, Stockton, CA 95211, USA
            \and
            Instituto de Astrof\'isica de Canarias, C/ Via L\'actea s/n, 38200 La Laguna, Canary Islands, Spain
            \and
            Departamento de Astrof\'isica, Universidad de La Laguna (ULL), E-38206 La Laguna, Tenerife, Spain 
            \and
            Mullard Space Science Laboratory, University College London, Holmbury St Mary, Dorking, Surrey RH5 6NT, UK
            }

   \date{}

   \abstract{Early-type dwarf galaxies are not simply featureless, old objects, but {were} found to be much more diverse, hosting substructures and a variety of stellar population properties. To explore the stellar content of {faint early-type galaxies}, and to investigate in particular those with recent central star formation, we study colours and colour gradients within one effective radius in optical ($g-r$) and near-infrared ($i-H$) bands for 120 Virgo cluster early-type galaxies with $-19~$mag $< M_{r} < -16~$mag. {Twelve galaxies turn out to have blue cores, when defined as $g-r$ colour gradients larger than 0.10~mag/$R_{\rm eff}$, which represents the positive tail of the gradient distribution. For these galaxies, we find that they have the strongest age gradients, and that even outside the blue core, their mean stellar population is younger than the mean of ordinary faint early-type galaxies.} The metallicity gradients of these blue-cored {early-type dwarf galaxies} are, however, in the range of most normal {faint early-type galaxies}, which we find to have non-zero gradients with higher central metallicity. The blue central regions are consistent with star formation activity within the last few 100 Myr. We discuss whether these galaxies could be explained by environmental quenching of star formation in the outer galaxy regions while the inner star formation activity continued.
}

   \maketitle

%-----------------------------------------------------------------------------------
%-----------------------------------------------------------------------------------
\section{Introduction}
\label{sec:intro}

{The most commonly occurring morphological type in galaxy clusters \citep{Binggeli1985} are early-type dwarf galaxies, defined as diffuse objects with smooth regular morphology \citep{SandageBinggeli1984}, which usually goes along with paucity of gas, dust, and star formation. At absolute $B$ magnitudes brighter than -14 mag and up to the end of the dwarf range at -18 mag ($10^8\lesssim M_{\rm stellar} \lesssim 10^{10}$ M$_\odot$), early-type dwarf galaxies have typically been classified as dwarf ellipticals (dE) and dwarf lenticulars \citep[dS0,][]{BinggeliCameron1991} or alternatively as spheroidals \citep[Sph,][]{Kormendy2009}, while they have been known as dwarf spheroidals (dSph) at fainter magnitudes \citep{Grebel2003}.} Their {abundance} and low mass means that they are ideal probes for studying the influence of high-density environments on galaxy formation and evolution. The Virgo cluster is the nearest high-density region and thus provides a perfect setting to investigate low-mass galaxies \citep[e.g.][]{Roberts2007,Boselli2008,Boselli2014,Toloba2015,Liu2016,Ferrarese2016}. Furthermore, Virgo is dynamically young and field galaxies and groups of galaxies are falling into the cluster \citep[e.g.][]{TullyShaya1984,Binggeli1987,Kim2016}. 

{While at first glance, most early-type dwarfs are small featureless objects, closer investigation has revealed that they can host substructures like weak spiral arms, bars, disks, nuclei, or blue cores \citep[e.g.][]{Jerjen2000,Barazza2002,DeRijcke2003,Lisker2006a,Lisker2006b,Lisker2007,Lisker2008,Janz2012}. These features have stirred the debate about their origin and evolution. Given the scarcity of early-type dwarfs in the field \citep{Binggeli1988}, most proposed mechanisms rely on the idea of environmental transformation of late types. These include galaxy harassment caused by} tidal forces from the overall cluster potential combined with occasional close high-speed encounters \citep{Moore1996,Moore1998,Smith2010,Smith2015}, as well as ram pressure stripping \citep{GunnGott1972,vanZee2004a} and galaxy starvation \citep{Larson1980}.

Since it is debatable whether high-mass and low-mass early-type galaxies follow common scaling relations, it remains unclear whether continuously varying contributions of the same internal and external processes are sufficient to understand the properties of present-day early-type galaxies across a wide mass range \citep[e.g.][]{GrahamGuzman2003,JanzLisker2008,Dabringhausen2008,Kormendy2009,Glass2011,DabringhausenKroupa2013,Lisker2013,Sybilska2017,Janz2017}. {Traditionally, faint early-type galaxies were separated into dwarf and normal early-types, with the term `dwarf' applied to the (visually) more diffuse galaxies \citep{SandageBinggeli1984}, which typically also have a shallower surface brightness profile \citep{BinggeliCameron1991}. At a given luminosity, this classification meant that a dwarf would have a {\it larger} effective radius than a non-dwarf galaxy. Since, however, the size-magnitude diagram does not show a gap between dwarfs and non-dwarfs, but instead an overlap, the SMAKCED\footnote[1]{Stellar content, MAss and Kinematics of Cluster Early-type Dwarfs, \url{http://www.smakced.net}} collaboration decided to target faint early-type galaxies regardless of their size and their previous classification as (non-)dwarf \citep{Janz2012,Janz2014}. Since our sample is based on SMAKCED, we follow the same approach as presented in their study in order to avoid any subjectivity in selecting galaxies. Furthermore, analyses of spectroscopy for small samples of galaxies show no dichotomy in the stellar population properties or in the angular momentum content between faint early-type galaxies classified previously as dwarfs and non-dwarfs \citep{Guerou2015,Toloba2014a}.}

{Spectroscopic stellar population analyses of galaxies are usually preferred over photometry, since the stellar population properties are encoded in a number of spectral features \citep{Vazdekis2012}. They are, however, often limited to small sample sizes and inner galaxy regions \citep[e.g.][]{Paudel2011,Toloba2014b}, due to the long exposure times required for spectroscopy of faint galaxies. Photometric studies investigating broad-band colours have the advantage of comprising larger numbers of objects and reaching larger radii even for more diffuse galaxies \citep[e.g.][]{DenBrok2011,Janz2014}. Although they offer less detailed information for individual galaxies, they are still able to reveal} information about the stellar ages and metallicities of galaxy populations \citep[e.g.][]{Schombert2016,Roediger2017}. {Therefore, they} provide a fundamental source for investigating their star formation and enrichment histories \citep[e.g.][]{Barazza2006,Kim2010} in connection with their local and large-scale environment \citep{Barazza2009,Lietzen2012,Poudel2016,Poudel2017}. Galaxy colour gradients contribute additional information \citep[see e.g.][]{Peletier1993,Pierini2002}, thereby helping to understand galaxy evolution. {However, \citet{Roediger2011} found no systematic dependence of the stellar population gradients of Virgo cluster galaxies on environment. For disk galaxies, they argue that those with positive age gradients may have had their outer gas disks stripped off, while those with negative age gradients have likely grown inside-out. For early-type dwarfs, \citeauthor{Roediger2011} pointed out that they seem incompatible with a primordial origin given their mostly positive age and negative metallicity gradients.

A particularly intriguing subclass of early-type dwarfs are those galaxies that possess young central stellar populations} \citep{Vigroux1984}. They have been found in nearly all environments, from virtual isolation \citep{Gu2006} to galaxy groups \citep{Hodge1973, TullyTrentham2008} and clusters \citep{DeRijcke2003,Lisker2006b}, in some cases even
dominating the early-type dwarf population \citep[Ursa Major cluster]{Pak2014}. Only in the massive Coma cluster were they found to be nearly absent \citep{DenBrok2011}. While \citet{Lisker2006b} {defined} the population of Virgo cluster blue-cored early-type dwarfs through visual inspection of $g-i$ colour maps, we present here a more quantitative analysis that makes use of the available optical and near-infrared imaging data.

This paper is organized as follows. In section \ref{sec:data} we give an overview of our sample. Our analysis is described in section \ref{sec:analysis}. In section \ref{sec:results} {the resulting} colours, colour gradients and their correlations are shown and separated into age- and metallicity-sensitive colours. The interpretation and discussion of our results in terms of stellar populations and galaxy evolution are presented in section \ref{sec:discuss}.

%-----------------------------------------------------------------------------------
%-----------------------------------------------------------------------------------
\section{Sample and data}
\label{sec:data}

We investigated a sample of 120 faint early-type galaxies in the Virgo cluster that span a luminosity range of $-19~$mag $< M_{r} < -16~$mag. Of these galaxies, 97 were classified as dwarf ellipticals and dwarf lenticulars and the other 23 galaxies were classified as normal (non-dwarf) ellipticals and lenticulars. This classification is based on the Virgo Cluster Catalog \citep[VCC,][]{Binggeli1985}, and takes into account the updated membership by heliocentric velocities from \citet{Lisker2013}. Our sample, based on \citet[][ SMAKCED]{Janz2014}, comprises 90\% of Virgo cluster early-type galaxies within $-19~$mag $< M_{r} < -17~$mag, and 50\% within $-17~$mag $< M_{r} < -16~$mag. {Within the faintest $\sim$0.5 mag of our sample's magnitude range, we are biased towards objects with brighter-than-average surface brightness. Our sample covers 14 of the 23 blue-cored early-type dwarf galaxies} that \citet{Lisker2006b} identified qualitatively from colour maps.

Optical $g$-, $r$-, and $i$-band images (background-subtracted, flux-calibrated, and astrometrically aligned) of our sample galaxies were provided by \citet{Lisker2007}, based on the imaging data of the {Sloan Digital Sky Survey (SDSS) Data Release 4 and Data Release 5} \citep{AdelmanMcCarthy2006,AdelmanMcCarthy2007}, with a pixel scale of 0.$\arcsec$396. Publicly available\footnote[2]{{\url{http://dc.zah.uni-heidelberg.de/browse/smakced/q}}} near-infrared $H$-band images (background-subtracted and flux-calibrated) from SMAKCED, obtained with the ESO New Technology Telescope (NTT), the Telescopio Nazionale Galileo, and the Nordic Optical Telescope, were provided by \citet{Janz2014}, with a pixel scale of 0.$\arcsec$25 and a median seeing of 0.$\arcsec$9. {Both data sets were corrected for Galactic extinction according to \citet{Schlegel1998}.}

In addition to the image treatment described in \citet{Lisker2007} and \citet{Janz2014} we had to align the $H$-band images with the optical images  {before measuring} colours and colour gradients. Therefore, we changed the pixel scale of the $H$-band images with IRAF\footnote[3]{Image Reduction and Analysis Facility, \url{http://iraf.noao.edu/}}/magnify to 0.$\arcsec$396 and aligned the $H$-band with IRAF/geomap and IRAF/geotran, feeding {these tasks} a list of objects in each image that was obtained with SExtractor \citep{BertinArnouts1996}\footnote[4]{\url{http://www.astromatic.net/software/sextractor}}, and visually inspecting the result. 

 In order to take into account differences in the {point spread functions (PSF)} of the different bands to the first order, we measured the {full width at half maximum (FWHM)} of several stars for each wavelength band with SExtractor and convolved the {respective} better image {of each pair of bands ($g$ \& $r$ and $i$ \& $H$)} with a Gaussian kernel with IRAF/gauss, using the quadratic difference of the PSF FWHM values and dividing it by 2.35 to yield the Gaussian~$\sigma$. {This yields a median FWHM in $g$ and $r$ of $1.\arcsec3$ with a $\pm 40\%$ range of [$1.\arcsec0$, $1.\arcsec6$], as well as a median FWHM in $i$ and $H$ of $1.\arcsec0$ and a $\pm 40\%$ range of [$0.\arcsec8$, $1.\arcsec3$]}.

%-----------------------------------------------------------------------------------
\section{Analysis}
\label{sec:analysis}

\begin{figure}[t!]
\includegraphics[width=87mm]{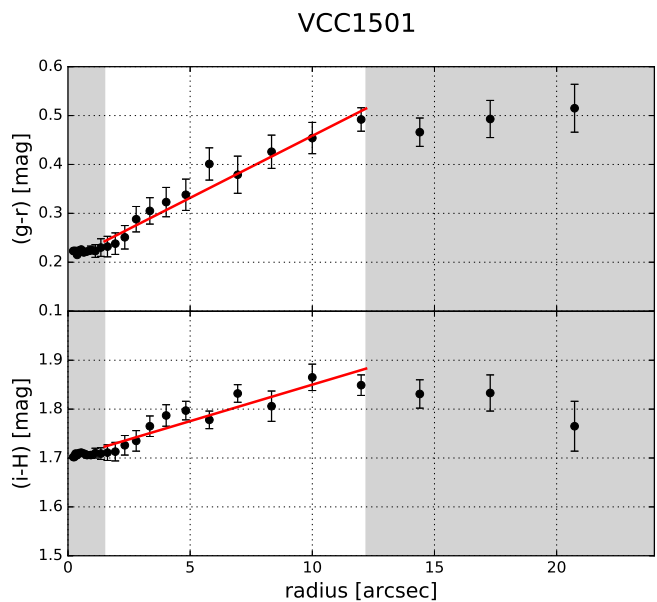}
\caption{{\bf Radial colour profile} in $g-r$ and $i-H$ for the galaxy VCC~1501. The $g-r$ colour (upper panel) and the $i-H$ colour (lower panel) are plotted against the radius of the galaxy. Linear gradients (red) {were fitted} from 1.$\arcsec$5 to the effective radius (12$\arcsec$), taking into account the uncertainties of the isophotal colours. {The resulting gradient values are} {$\delta (g-r) = 0.305 \pm 0.027$ mag/$R_{\rm eff}$ and $\delta (i-H) = 0.179 \pm 0.023$ mag/$R_{\rm eff}$.} {The area not used for the fit is shaded in grey.}}
\label{fig:radial-profile-vcc1501}
\end{figure}

 To extract radial colour profiles in $g-r$ and $i-H$ we measured the average intensity in each band in about 20-30 elliptical annuli for each galaxy with IRAF/ellipse between the centre and the two effective radii. {We use the semi-major axis of the half-light aperture in $H$ (published by \citealt{Janz2014}\footnotemark[2]) as effective radius.} The semi-major axis of {the elliptical annuli} increases by 20\% from one to the next. While running IRAF/ellipse we kept fixed the centre, the position angle and the ellipticity (measured in $r$ at the effective radius) of the galaxy. This defines a radial colour profile for each galaxy in both $g-r$ and $i-H$. We then fitted a weighted linear regression to the radial profiles using the non-linear least-squares Marquardt-Levenberg algorithm and taking into account the uncertainties of the isophotal colours with a weight of 1/$\sigma^{2}$, where $\sigma$ is the standard deviation of the respective colour value. We limited the range of our fits to radial bins with centres from 1.$\arcsec$5 up to {one} effective radius since the median seeing of the SDSS $r$-band images is 1.$\arcsec$4 \citep{AdelmanMcCarthy2006,AdelmanMcCarthy2007}. We decided not to use the innermost colour values despite our approximate homogenization of the PSFs of different bands. We made this decision because the specific PSF shape at the position of the galaxy centre may not be accurately reflected by a simple Gaussian; in addition, stellar nuclei can lead to rather steep central intensity profiles, thus affecting central galaxy colours in case of PSF differences. The outer limit of our fit is set by the desire for a high signal-to-noise ratio, i.e. small statistical uncertainties, and by a visual inspection of the colour profiles which shows that profiles beyond the effective radius are mostly flat or very shallow (consistent with \citealt{Lisker2006b}), whereas the main interest of our study are objects with strong gradients. 

\begin{figure}[t!]
\includegraphics[width=87mm]{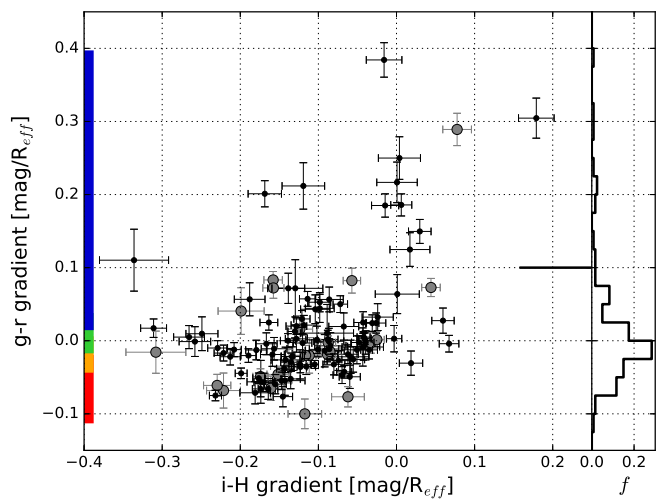}
\caption{Fitted $i-H$ gradients plotted against the fitted $g-r$ gradients. {The error bars denote the standard errors (see text).} Galaxies with $g-r$ gradients larger than 0.10 mag/$R_{\rm eff}$ are defined as quantitative blue-cored early-type dwarf galaxies and are located above the black horizontal tick in the plot. {Galaxies that were classified as non-dwarfs by \citet{Binggeli1985} are displayed as filled grey circles.} The colour-coding on the y-axis defines the ranges of four subsamples that all include the same number of galaxies (30). The fraction $f$ of galaxies in the $g-r$ gradient distribution is presented in the right histogram.}
\label{fig:gradients-gr-iH}
\end{figure}

 This gives us quantitative radial colour gradients, defined by their slopes d($g-r$)/d($R/R_{\rm eff}$) and d($i-H$)/d($R/R_{\rm eff}$), and parameter errors achieved by the fit, which are calculated as standard errors of a linear least-squares problem. {In addition to these fitting uncertainties, we quantify the goodness of fit with the coefficient of determination, $\mathcal{R}^{2}$, on a scale from 0 to 1 (the larger, the better). The value of $\mathcal{R}^{2}$ is provided in Table 1 and helps to indicate how well our chosen linear fit is able to approximate a galaxy's colour profile. An example is provided in Fig.~\ref{fig:radial-profile-vcc1501}, which shows the $g-r$ and $i-H$ radial profile of the galaxy VCC~1501 and its fitted linear slopes from 1.$\arcsec$5 to the effective radius, 12$\arcsec$. This galaxy has one of the most extreme colour gradients in our sample: $\delta (g-r) = 0.305 \pm 0.027$ mag/$R_{\rm eff}$ ($\mathcal{R}^{2}_{\delta (g-r)} = 0.95$) and $\delta (i-H) = 0.179 \pm 0.023$ mag/$R_{\rm eff}$ ($\mathcal{R}^{2}_{\delta (i-H)} = 0.85$).}

Since the existence of colour gradients complicates the definition of a global galaxy colour, we decided to choose the colour at the effective radius as representative of the main body, which avoids the contamination due to blue centres if we were to use {radially} integrated colours. {In order to take the colour-magnitude relation of early-type dwarfs \citep{Lisker2008} into account, we computed residual colours as follows. For a given galaxy with absolute magnitude $M_{r}$, we calculated the mean colour of all galaxies that lie in a magnitude interval of $M_{r} \pm 0.5~$mag. The difference between the galaxy's colour value and the mean colour is its residual colour. For example, galaxy VCC~1488 has a $g-r$ colour of $0.52~$mag at the effective radius and an absolute magnitude of $M_{r} = -17.0$~mag. The mean colour $\left<(g-r)\right>$ of all galaxies in the magnitude interval $-17.5~$mag $< M_{r} < -16.5~$mag is $0.62~$mag and hence the residual $g-r$ colour of VCC~1488 is $-0.10~$mag. %\emph{When we talk about colours in the further course of this work we will always refer to this definition of residual colours.}
}

%-----------------------------------------------------------------------------------
%-----------------------------------------------------------------------------------
\section{Results}
\label{sec:results}
\subsection{Colours and colour gradients}
\label{subsec:col-colgrad}

Our analysis yields linear $g-r$ and $i-H$ gradients for 120 faint early-type galaxies in the Virgo cluster {(Fig.~\ref{fig:gradients-gr-iH}). There is no systematic difference between the colour gradients of galaxies originally classified as dwarfs and as non-dwarfs in \citet[][also see Sect.~\ref{sec:intro}]{Binggeli1985}.}

\begin{table*}
\caption{{Properties of quantitative blue-cored early-type dwarf galaxies: (1) designation, (2) absolute $r$-band magnitude, (3) effective radius (semi-major axis of half-light aperture in $H$), (4) $g-r$ gradient fitted from 1.$\arcsec$5 to $R_{\rm eff}$, (5) coefficient of determination for $\delta$($g-r$), (6) $i-H$ gradient fitted from 1.$\arcsec$5 to $R_{\rm eff}$, (7) coefficient of determination for $\delta$($g-r$), (8) $g-r$ colour measured at 1.$\arcsec$5, (9) $i-H$ colour measured at 1.$\arcsec$5, (10) $g-r$ colour measured at $R_{\rm eff}$, (11) $i-H$ colour measured at $R_{\rm eff}$}.} % title of Table
\label{tab:dEbcs}      % is used to refer this table in the text
\centering                          % used for centering table
\begin{tabular}{c c c c c c c c c c c}        % centered columns (4 columns)
\hline\hline  \\               % inserts double horizontal lines
Object & $M_{r}$ & $R_{\rm eff}$ & $\delta$($g-r$) & $\mathcal{R}^{2}_{\delta(g-r)}$ & $\delta$($i-H$) & $\mathcal{R}^{2}_{\delta(i-H)}$ & ($g-r$)$_{1.\arcsec5}$ & ($i-H$)$_{1.\arcsec5}$ & ($g-r$)$_{R_{\rm eff}}$ & ($i-H$)$_{R_{\rm eff}}$   \\ 
VCC &   [mag] & [arcsec] & [mag/$R_{\rm eff}$] & & [mag/$R_{\rm eff}$] & & [mag] & [mag] & [mag] & [mag] \\% table heading 
 \\ \hline    \hfill \\          % inserts single horizontal line
\phantom{0}0021 & $-$ 17.4 & 14.2 &  0.22 $\pm$0.03 & 0.90 & \phantom{$-$} 0.00 $\pm$0.03 & 0.00 & 0.33 $\pm$0.02  &  1.83 $\pm$0.02  &  0.50 $\pm$0.03  &  1.83 $\pm$0.02\\
\phantom{0}0173 & $-$ 16.9 & 14.5 &  0.15 $\pm$0.02 & 0.45 & \phantom{$-$} 0.03 $\pm$0.01 & 0.24 & 0.39 $\pm$0.02  &  1.79 $\pm$0.02  &  0.57 $\pm$0.02  &  1.82 $\pm$0.01\\
\phantom{0}0209 & $-$ 16.9 & 16.9 &  0.11 $\pm$0.04 & 0.88 & $-$ 0.34 $\pm$0.04 & 0.88 & 0.53 $\pm$0.06  &  2.06 $\pm$0.06  &  0.62 $\pm$0.04  &  1.81 $\pm$0.06\\
\phantom{0}0870 & $-$ 17.0 & 16.7 &  0.25 $\pm$0.03 & 0.91 & \phantom{$-$} 0.00 $\pm$0.03 & 0.02 & 0.36 $\pm$0.04  &  1.81 $\pm$0.04  &  0.57 $\pm$0.03  &  1.82 $\pm$0.03\\
\phantom{0}0951 & $-$ 17.8 & 19.3 &  0.19 $\pm$0.02 & 0.88 & $-$ 0.02 $\pm$0.02 & 0.03 & 0.45 $\pm$0.02  &  1.83 $\pm$0.02  &  0.58 $\pm$0.02  &  1.78 $\pm$0.02\\
\phantom{0}1488 & $-$ 17.0 & 14.0 &  0.19 $\pm$0.02 & 0.84 & \phantom{$-$} 0.01 $\pm$0.01 & 0.01 & 0.37 $\pm$0.01  &  1.88 $\pm$0.01  &  0.52 $\pm$0.02  &  1.88 $\pm$0.02\\
\phantom{0}1499 & $-$ 16.8 &  7.5 &  0.29 $\pm$0.02 & 0.99 & \phantom{$-$} 0.08 $\pm$0.02 & 0.68 & 0.16 $\pm$0.01  &  1.65 $\pm$0.01  &  0.39 $\pm$0.02  &  1.71 $\pm$0.02\\
\phantom{0}1501 & $-$ 16.3 & 12.0 &  0.31 $\pm$0.03 & 0.95 & \phantom{$-$} 0.18 $\pm$0.02 & 0.85 & 0.23 $\pm$0.02  &  1.71 $\pm$0.02  &  0.49 $\pm$0.02  &  1.85 $\pm$0.02\\
\phantom{0}1512 & $-$ 16.5 & 12.1 &  0.21 $\pm$0.03 & 0.52 & $-$ 0.12 $\pm$0.03 & 0.82 & 0.40 $\pm$0.04  &  1.94 $\pm$0.04  &  0.67 $\pm$0.03  &  1.83 $\pm$0.02\\
\phantom{0}1684 & $-$ 16.7 & 18.5 &  0.38 $\pm$0.02 & 0.88 & $-$ 0.02 $\pm$0.02 & 0.05 & 0.20 $\pm$0.02  &  1.77 $\pm$0.02  &  0.50 $\pm$0.02  &  1.75 $\pm$0.02\\
\phantom{0}1779 & $-$ 16.9 & 19.6 &  0.20 $\pm$0.02 & 0.76 & $-$ 0.17 $\pm$0.02 & 0.92 & 0.36 $\pm$0.02  &  1.86 $\pm$0.03  &  0.55 $\pm$0.02  &  1.69 $\pm$0.02\\
\phantom{0}1912 & $-$ 17.9 & 22.2 &  0.13 $\pm$0.02 & 0.86 & \phantom{$-$} 0.02 $\pm$0.03 & 0.21 & 0.46 $\pm$0.09  &  1.90 $\pm$0.10  &  0.60 $\pm$0.01  &  1.93 $\pm$0.02\\ \hfill \\
\hline                                   %inserts single line
\end{tabular}
%\tablefoot{}
\end{table*}

\begin{figure*}
{\bf A: Quantitative blue-cored early-type dwarf galaxies} \\  
\includegraphics[width=0.245\textwidth]{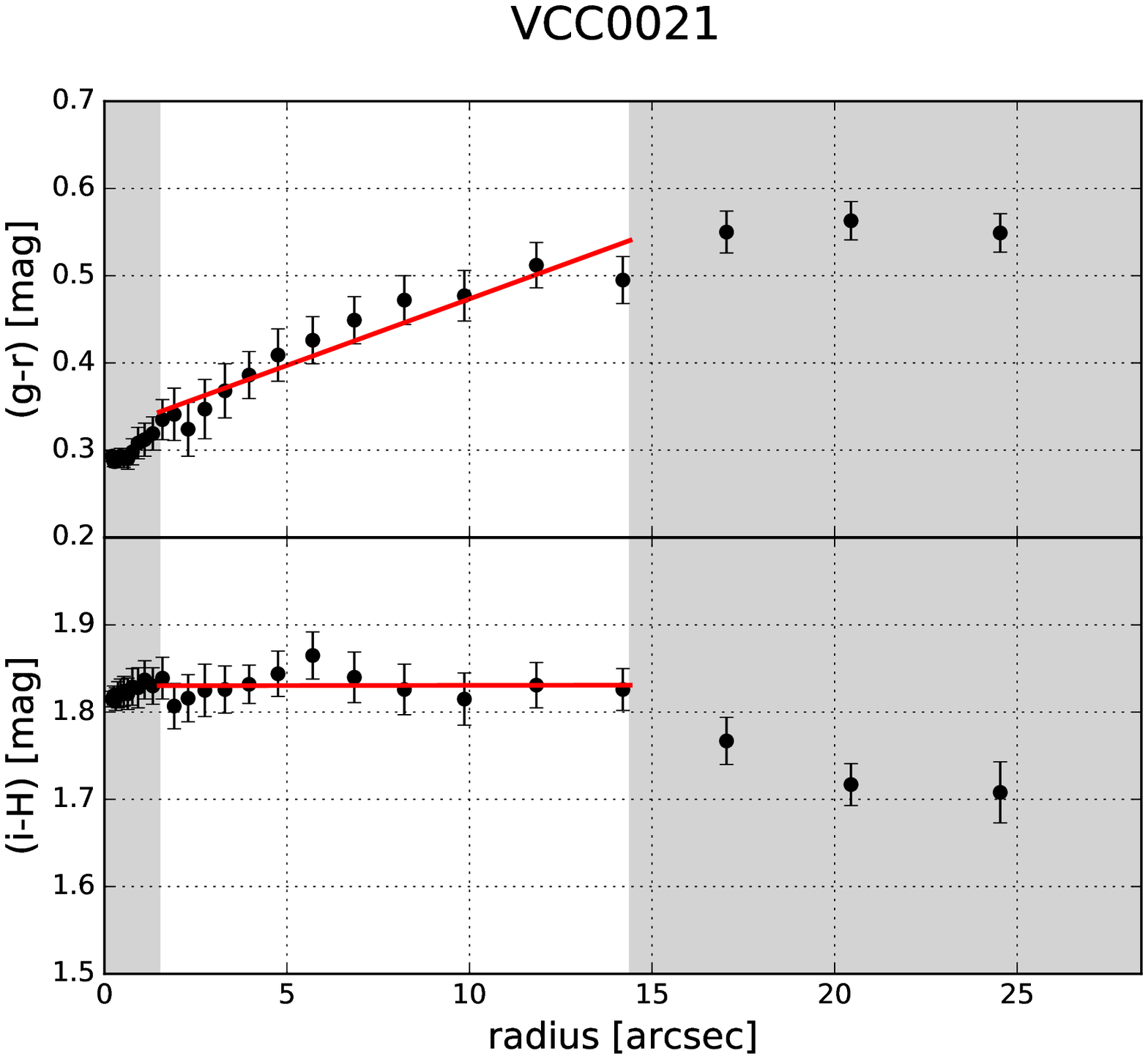}
\includegraphics[width=0.245\textwidth]{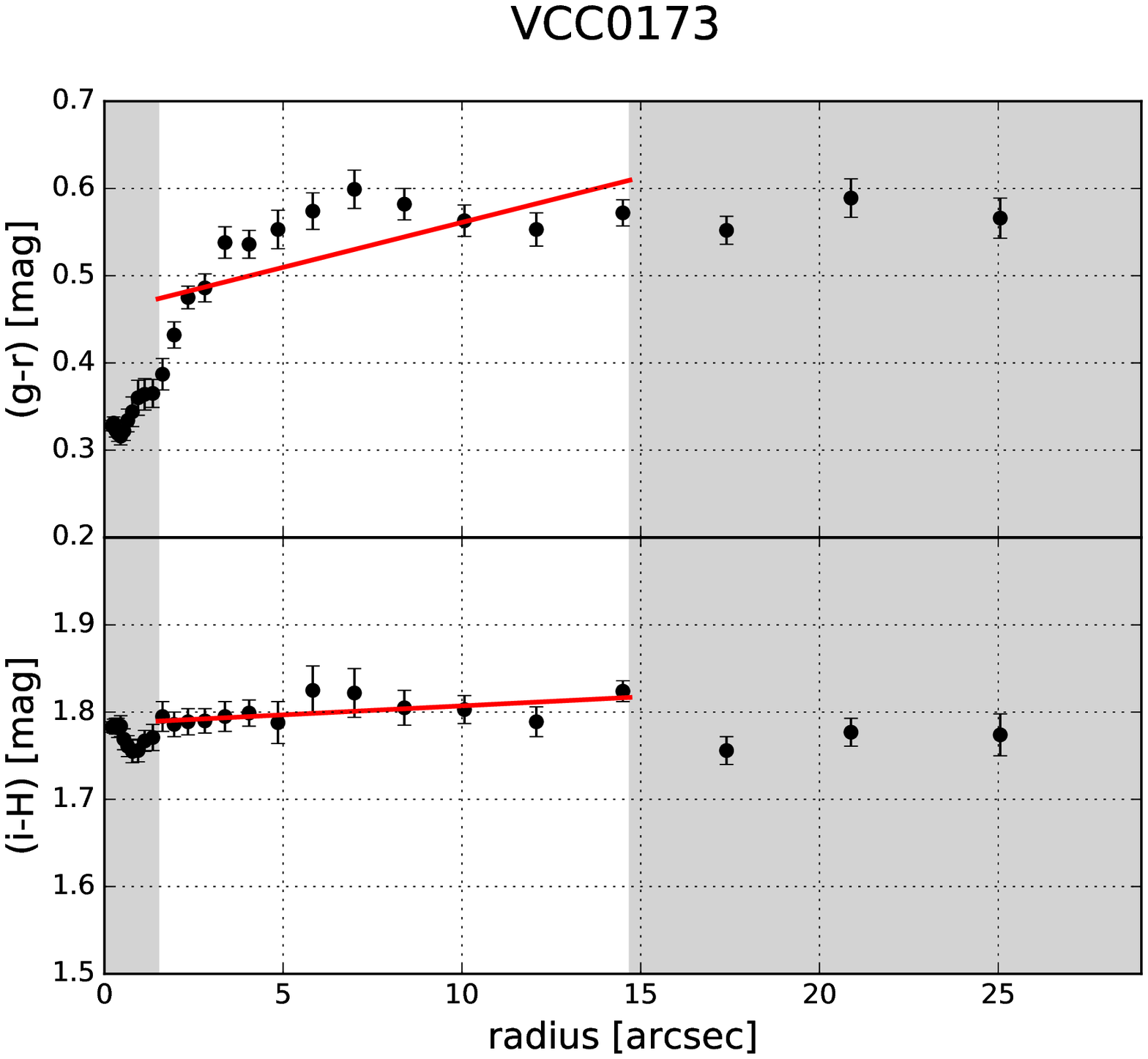}
\includegraphics[width=0.245\textwidth]{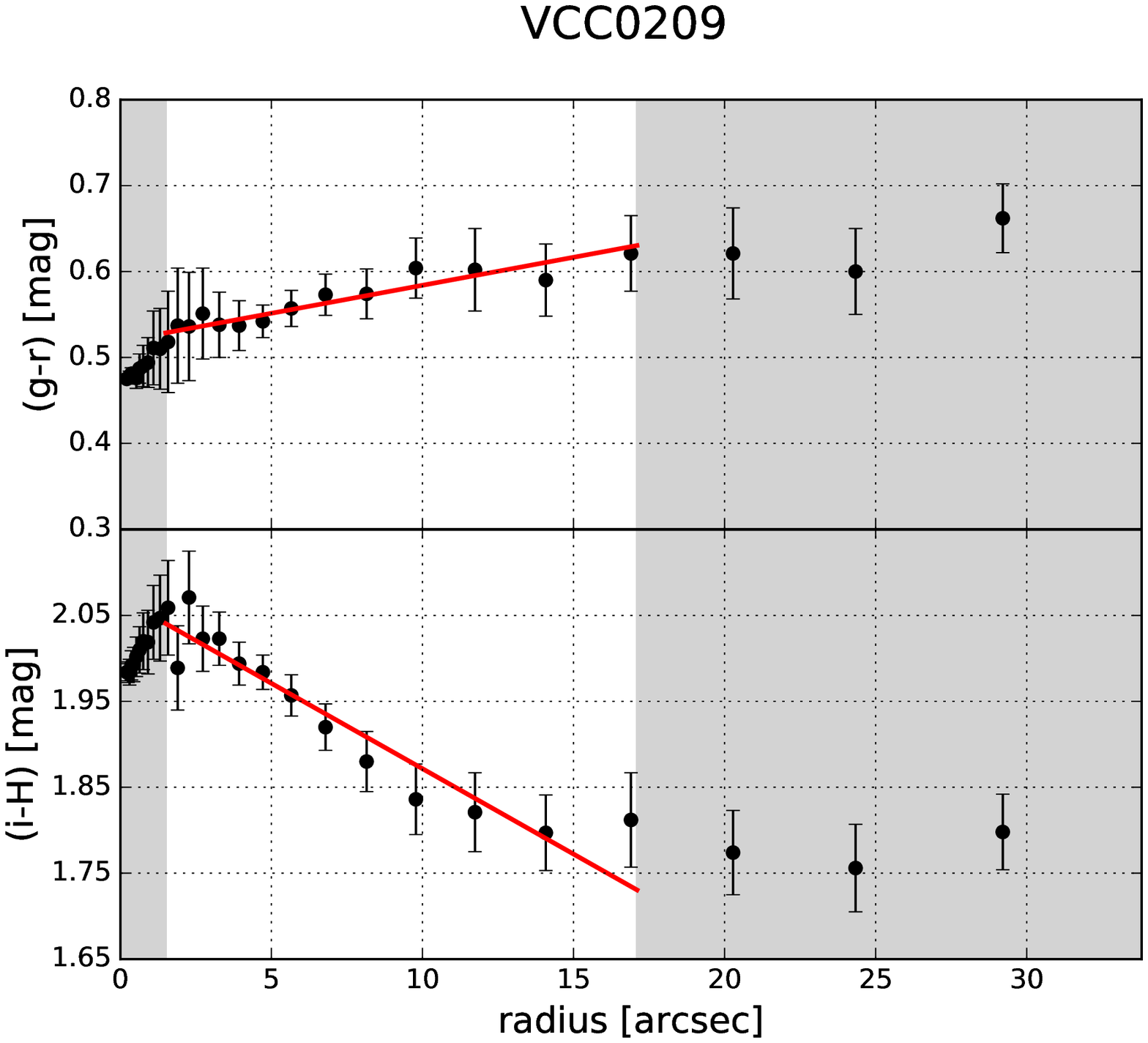} 
\includegraphics[width=0.245\textwidth]{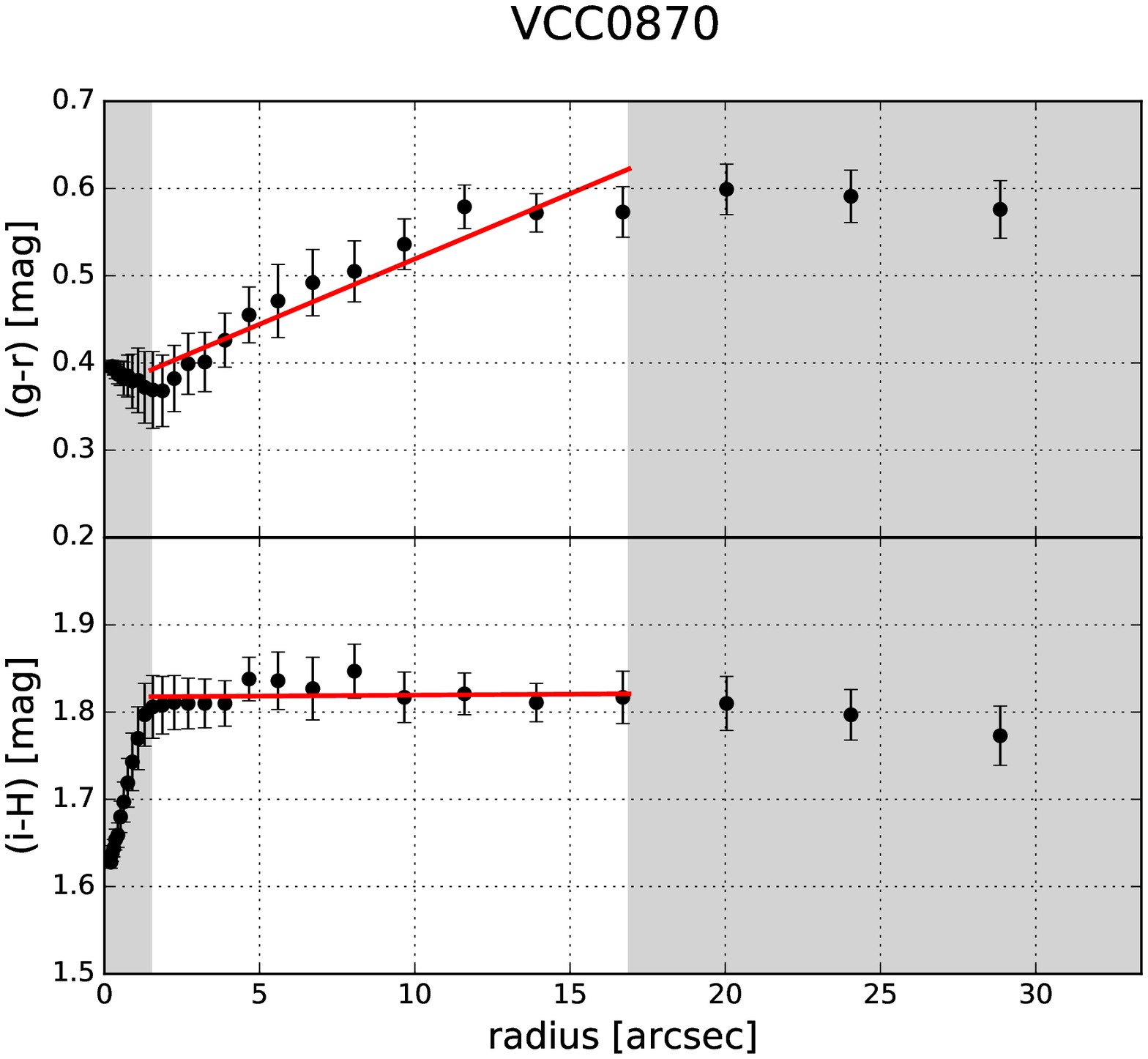} \\
\includegraphics[width=0.245\textwidth]{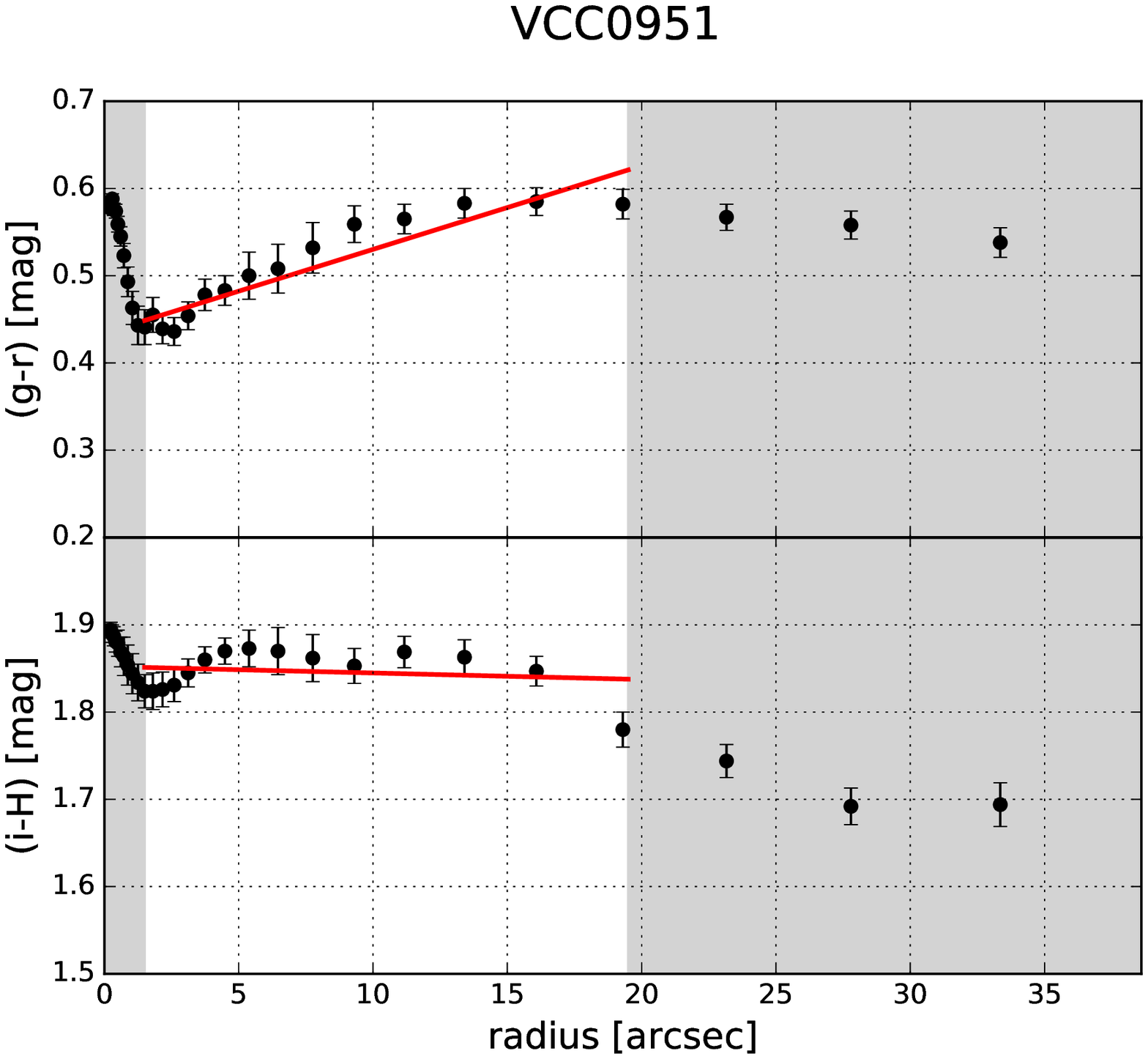} 
\includegraphics[width=0.245\textwidth]{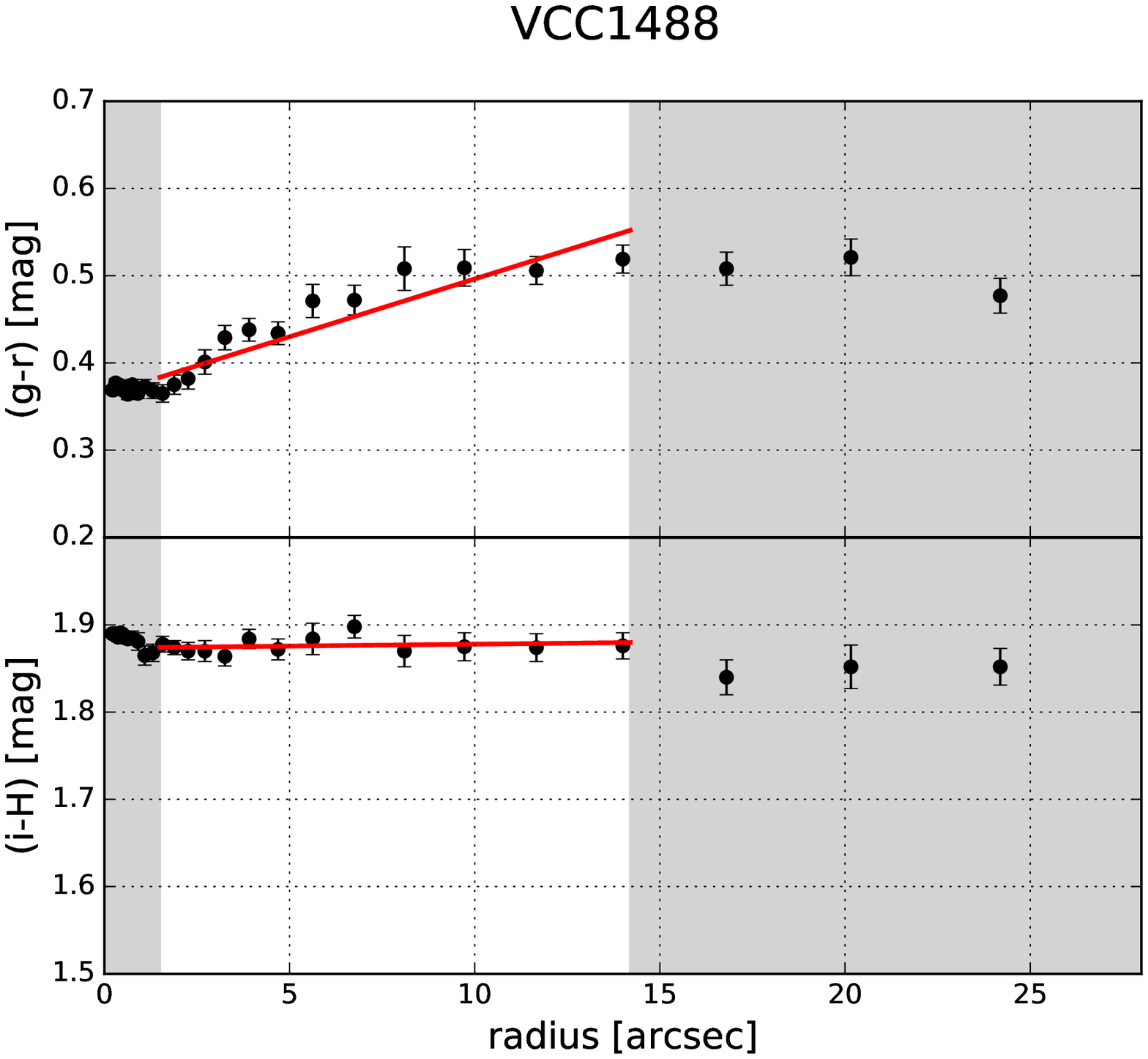}
\includegraphics[width=0.245\textwidth]{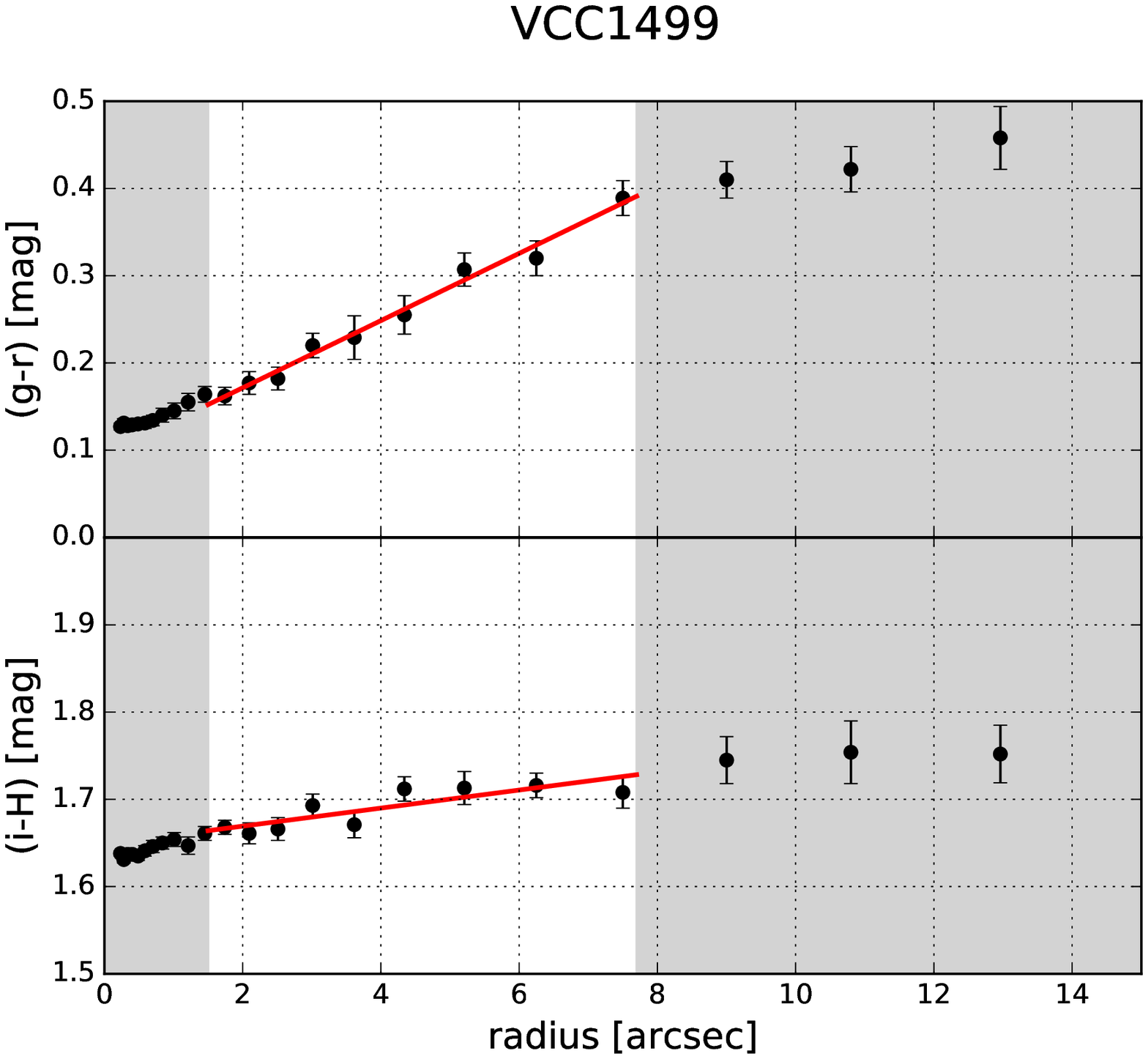}
\includegraphics[width=0.245\textwidth]{radial1501_g-r__i-H.eps} \\
\includegraphics[width=0.245\textwidth]{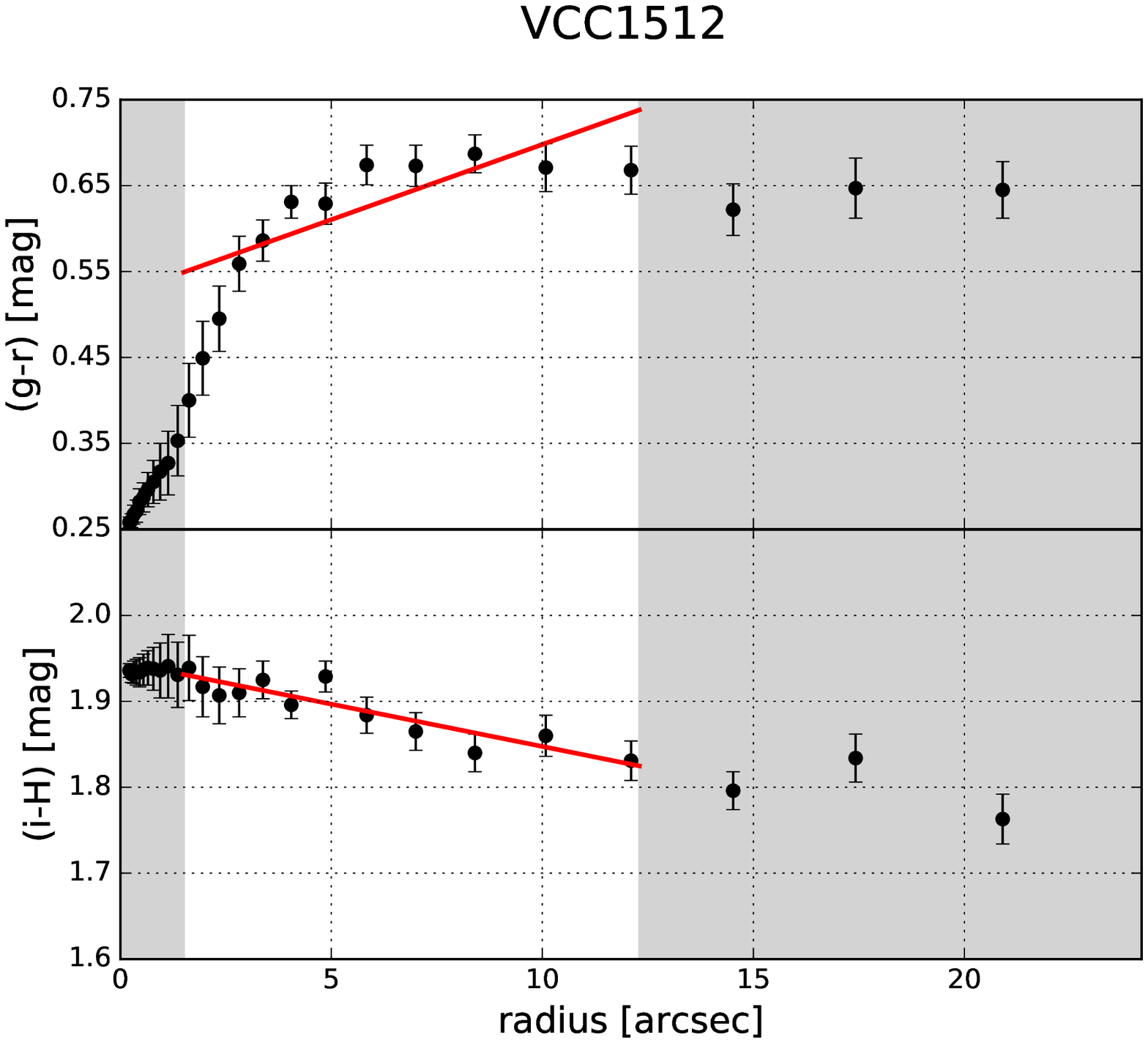} 
\includegraphics[width=0.245\textwidth]{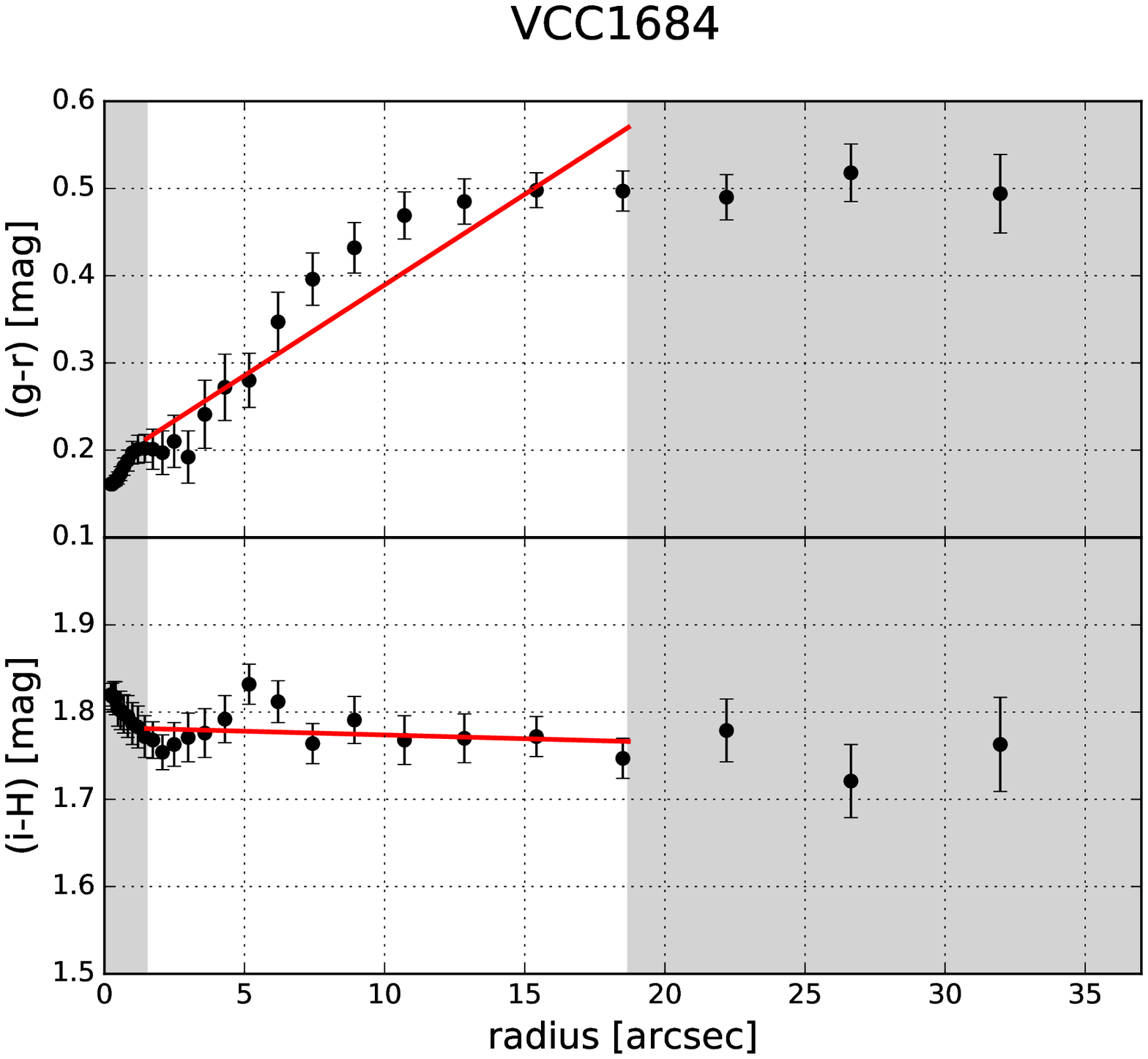}
\includegraphics[width=0.245\textwidth]{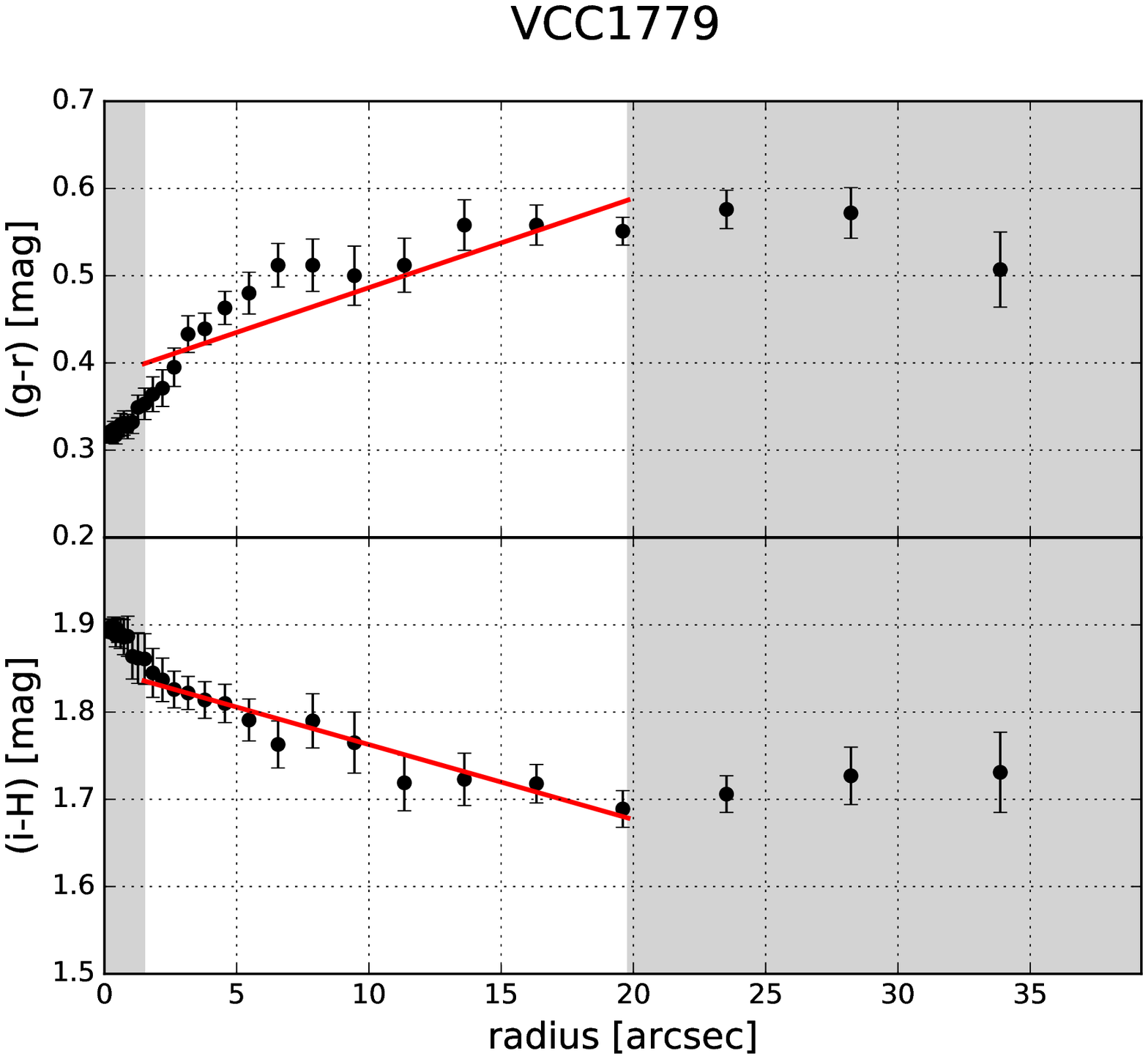}
\includegraphics[width=0.245\textwidth]{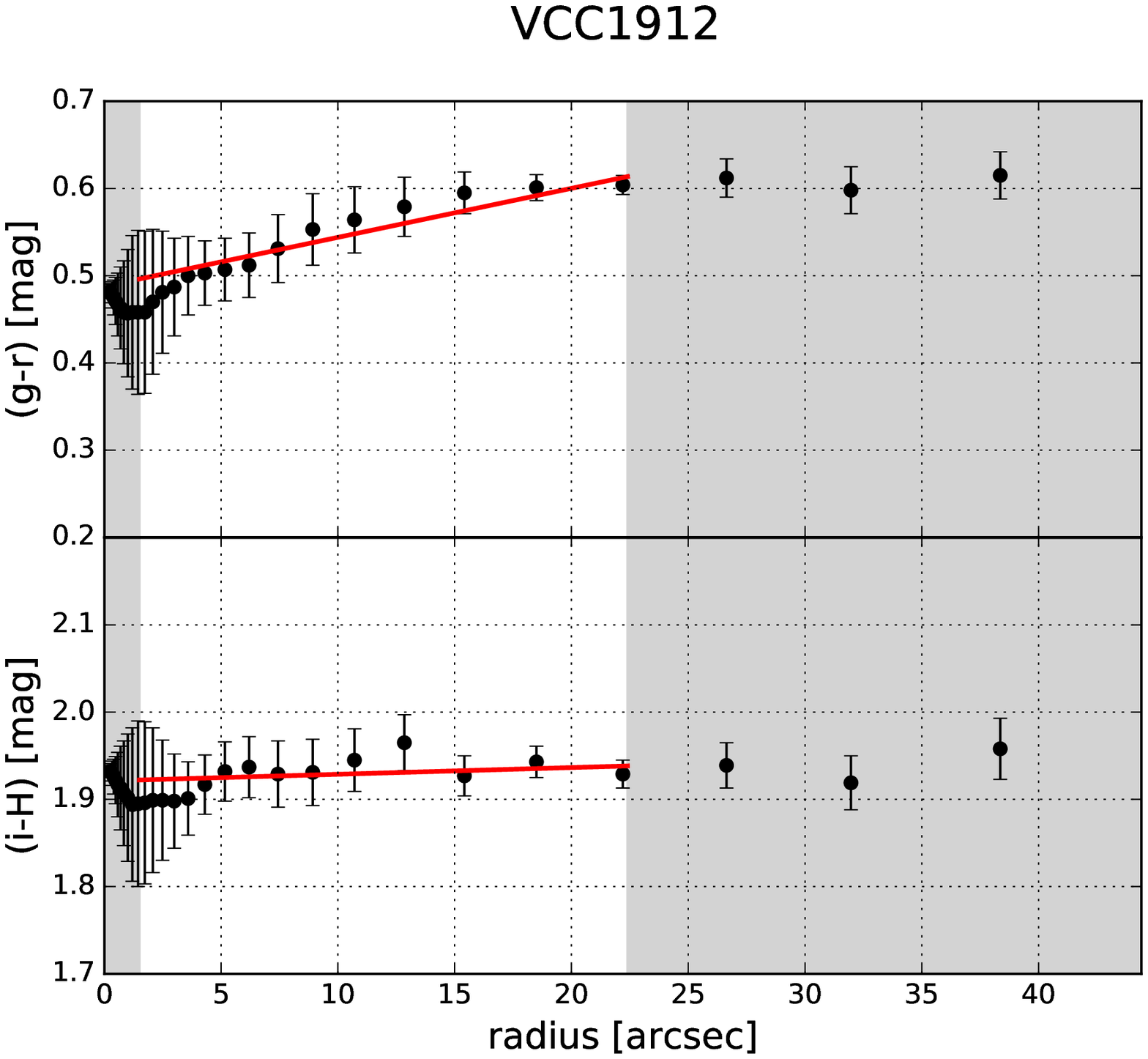} \\ \hfill  \\  
{\bf B: No quantitative blue-cored early-type dwarf galaxies} \\ 
\includegraphics[width=0.245\textwidth]{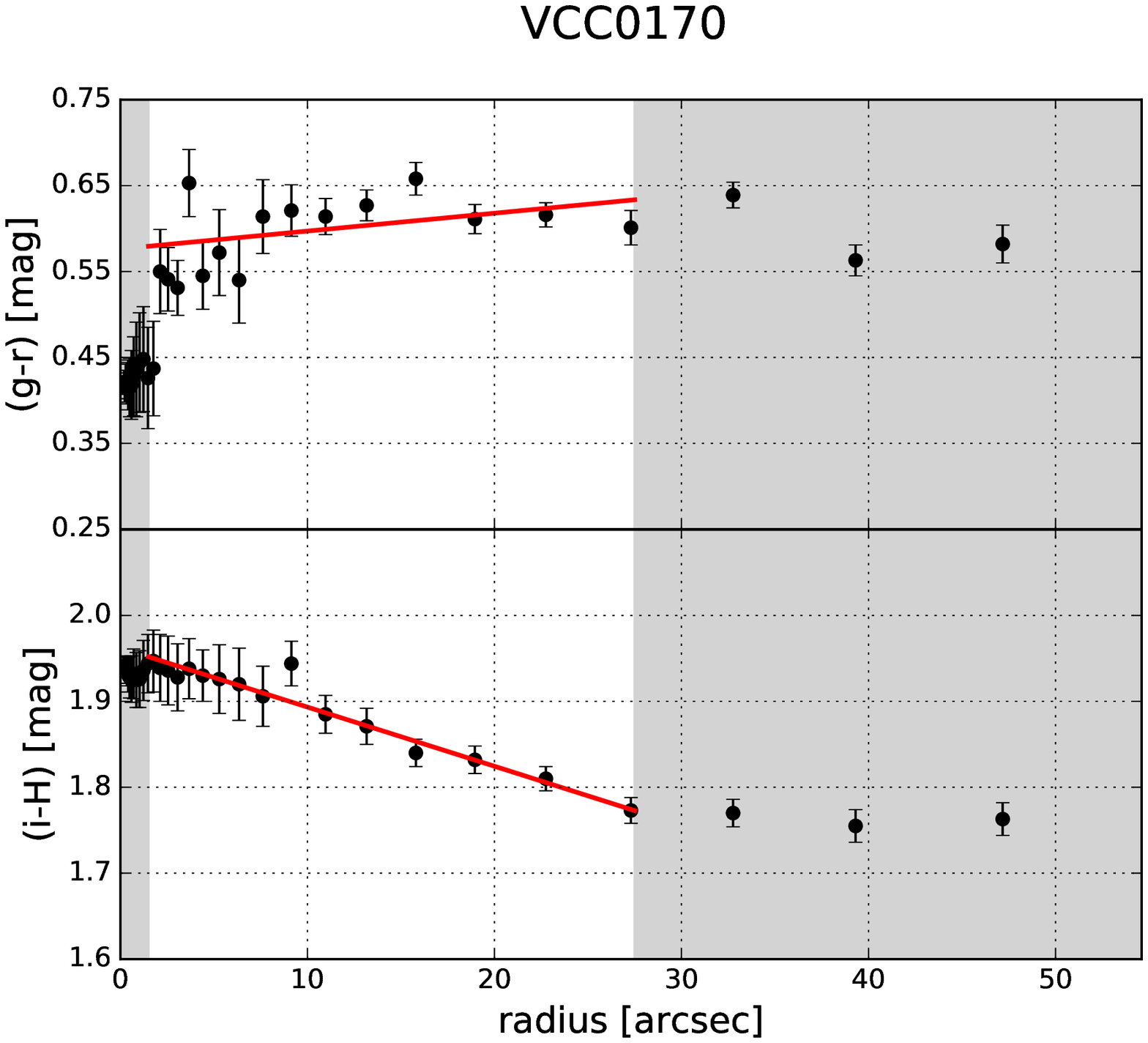}
\includegraphics[width=0.245\textwidth]{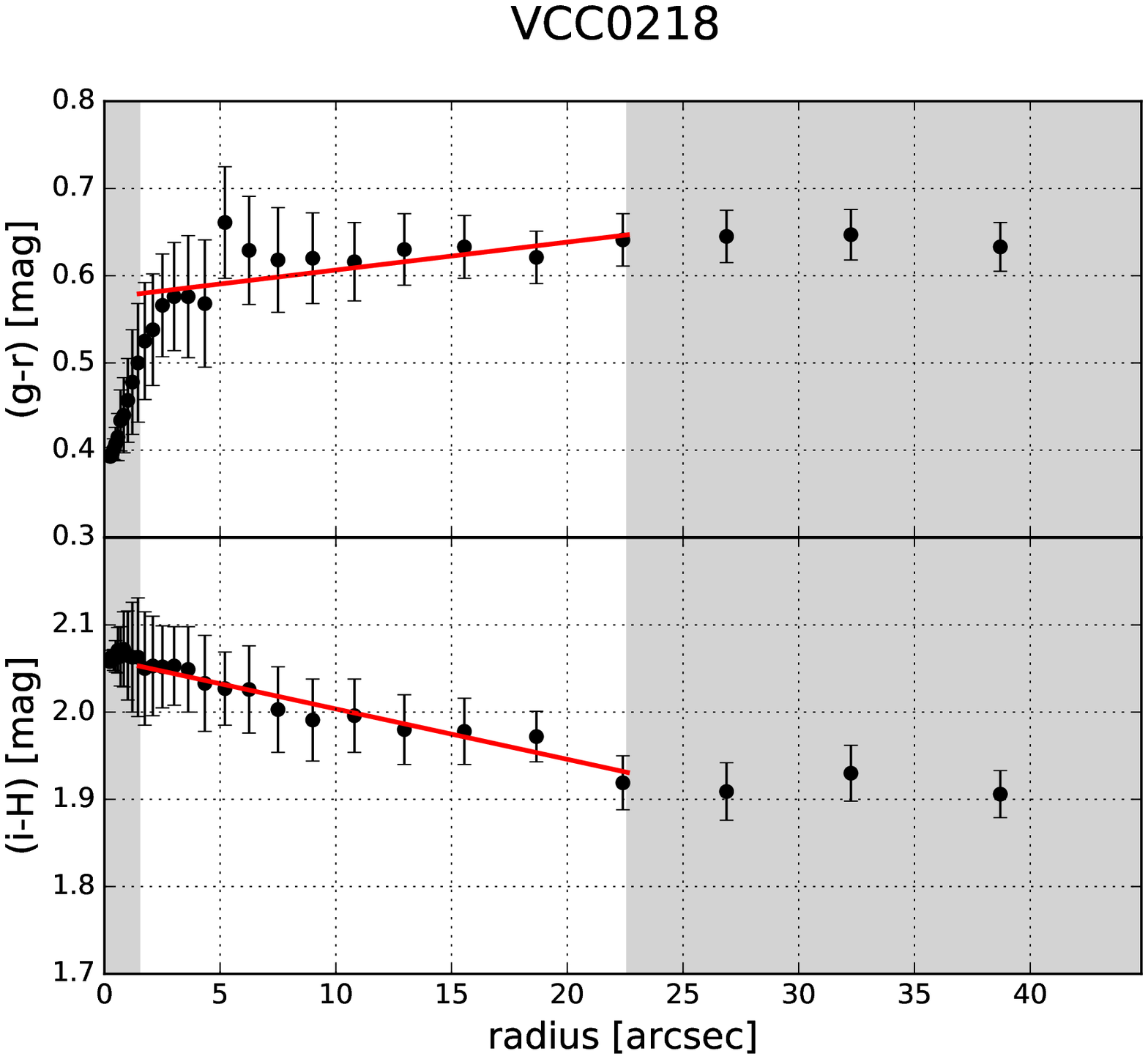} 
\includegraphics[width=0.245\textwidth]{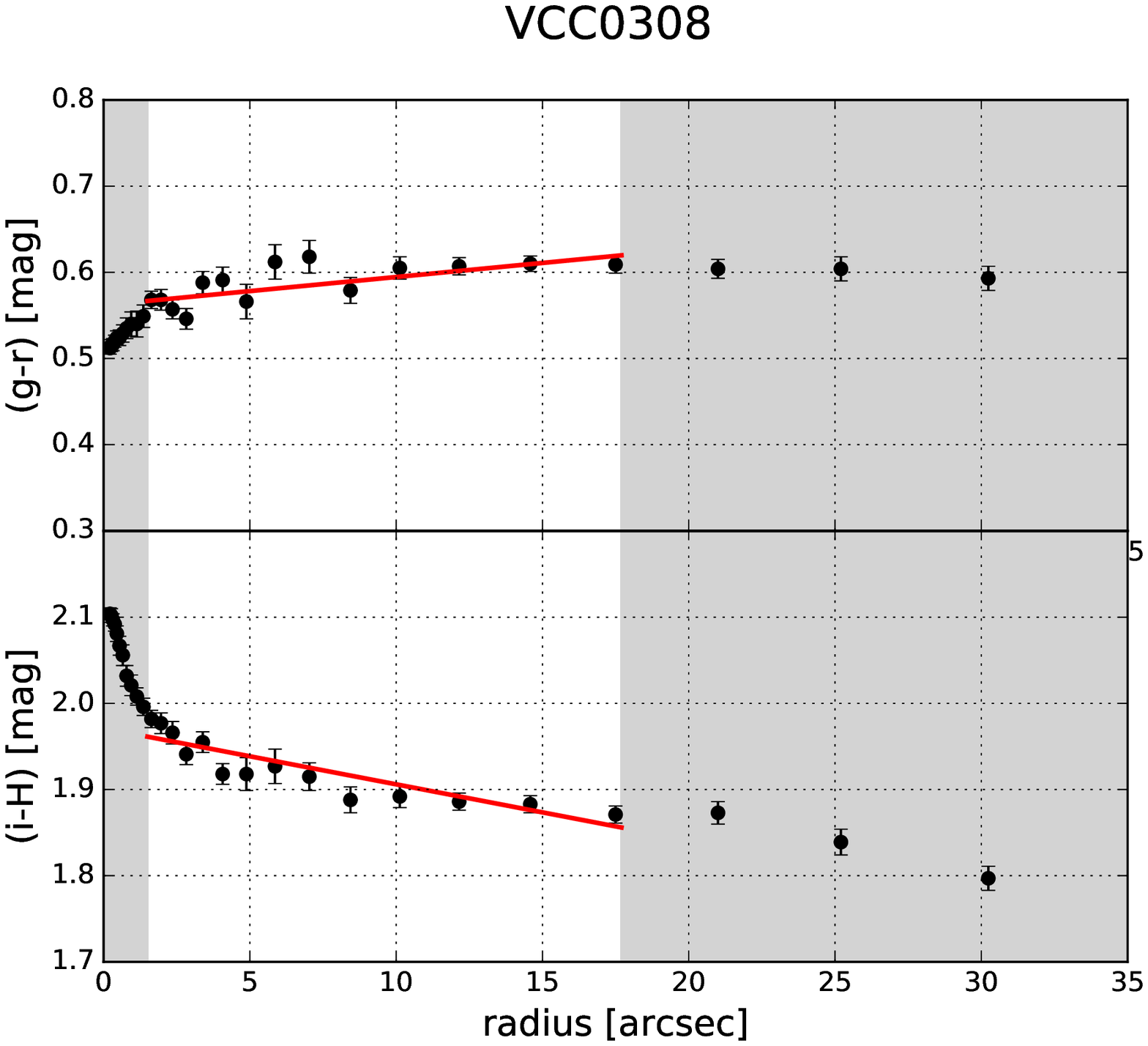}
\includegraphics[width=0.245\textwidth]{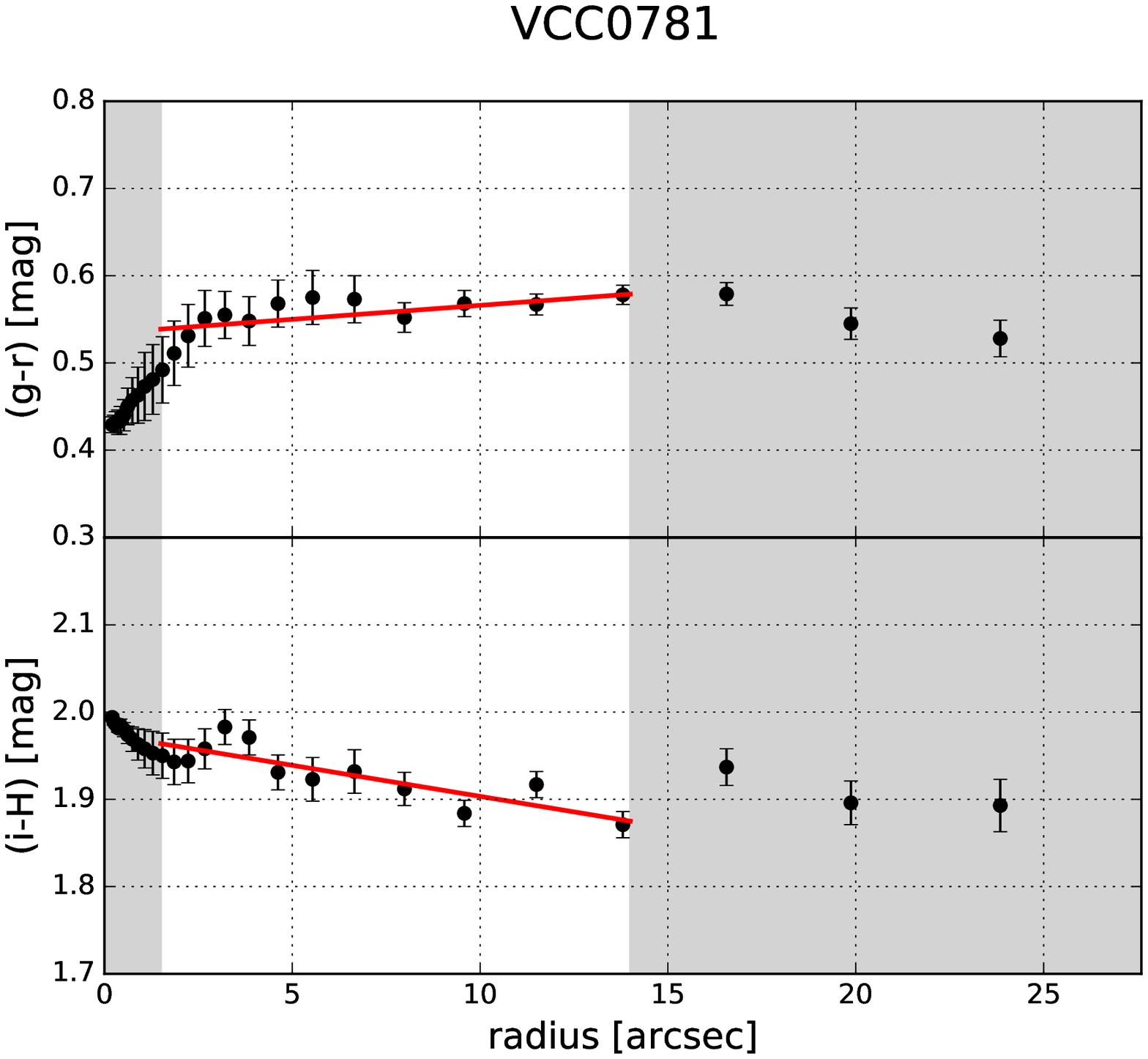} \\ \hfill  \\ 
{\bf C: Normal early-type dwarf galaxies} \\ 
\includegraphics[width=0.245\textwidth]{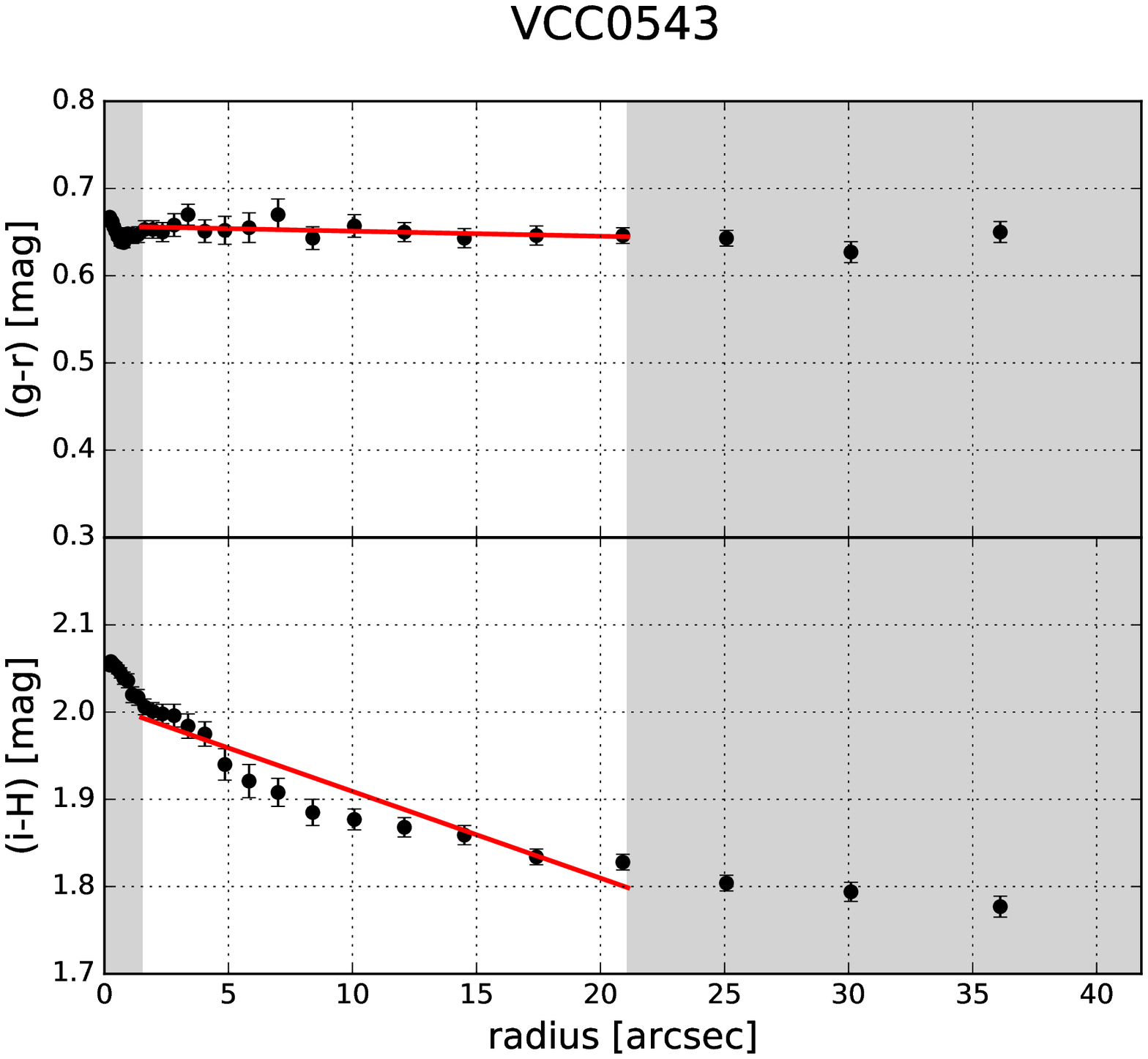}
\includegraphics[width=0.245\textwidth]{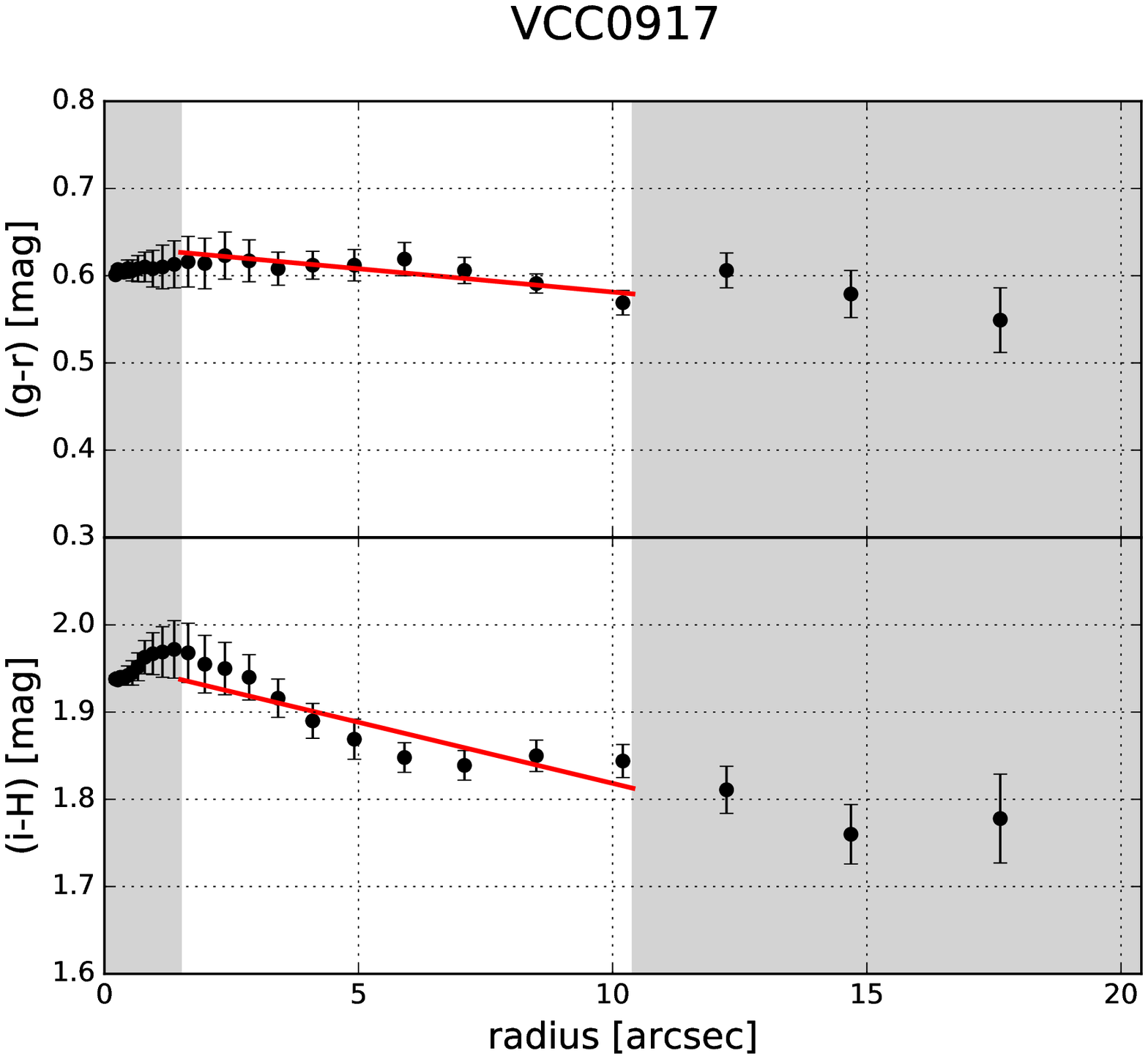}
\includegraphics[width=0.245\textwidth]{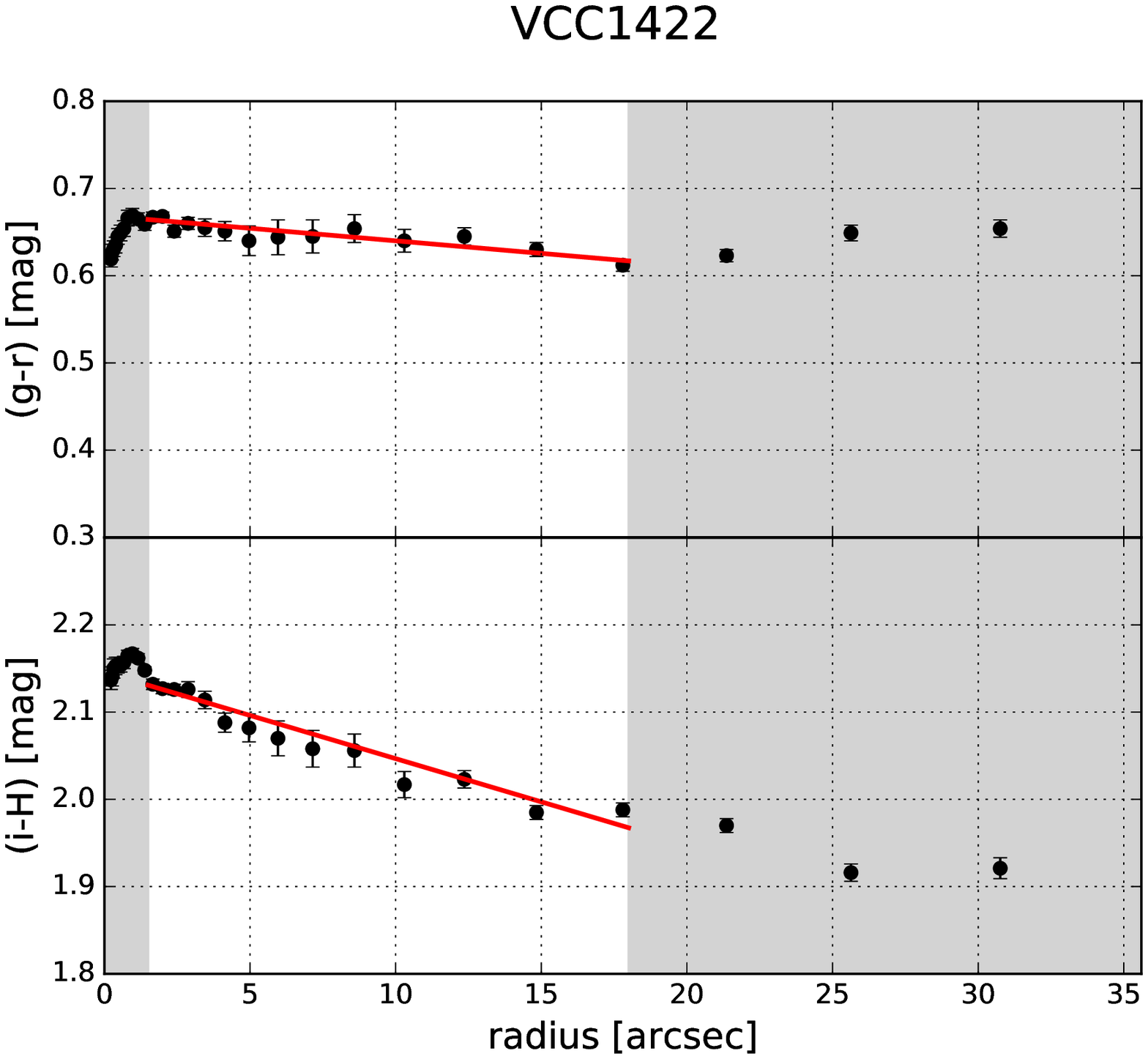}
\includegraphics[width=0.245\textwidth]{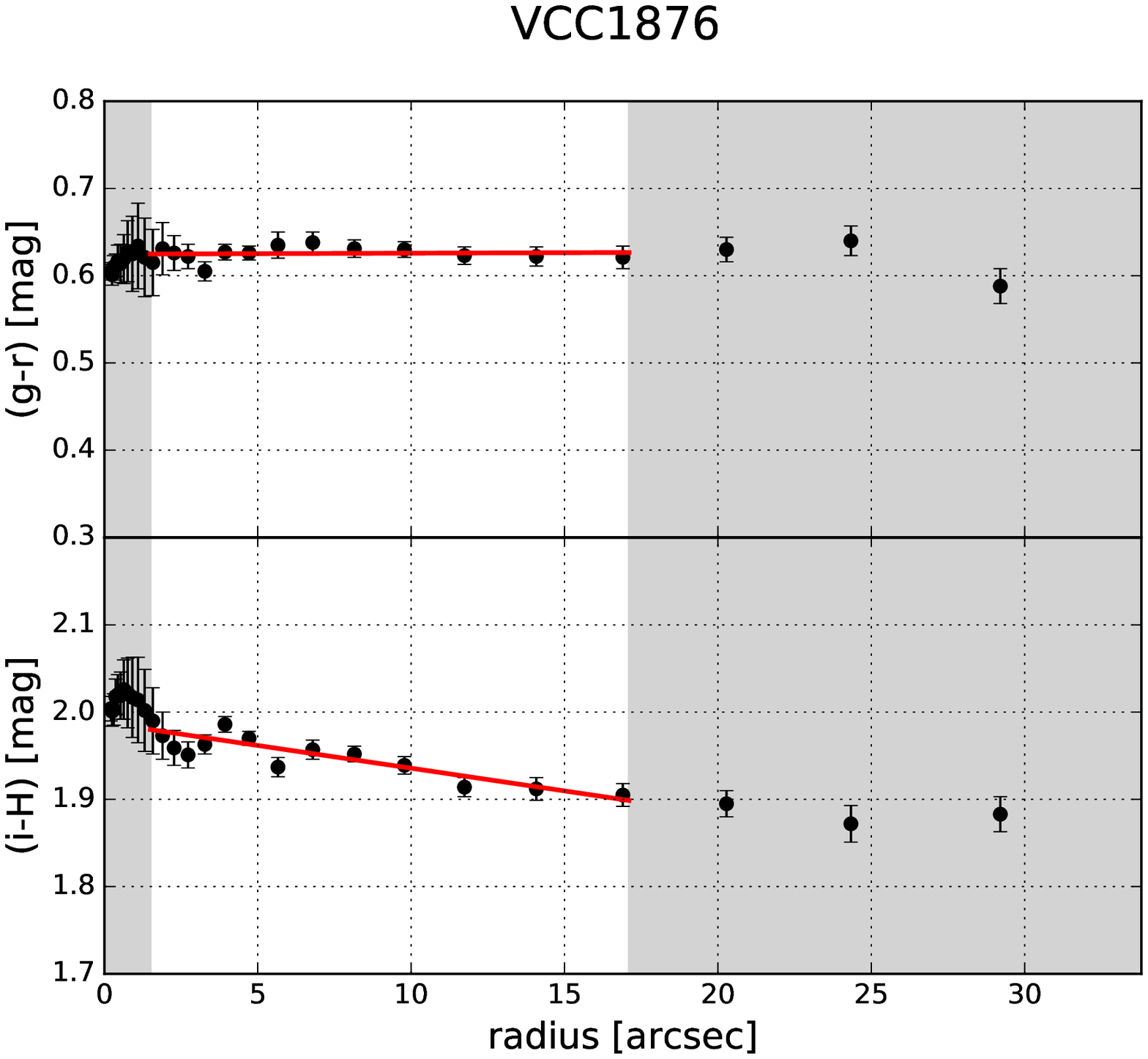} \\
\caption{{{\bf A:} Radial colour profiles in $g-r$ and $i-H$ (same as in Fig.~\ref{fig:radial-profile-vcc1501}) for all 12 quantitative blue-cored ($g-r$ gradient $> 0.10$ mag/$R_{\rm eff}$) early-type dwarf galaxies. {\bf B:} Radial colour profiles of the four galaxies that were classified as qualitative blue-cored early-type dwarf galaxies by \citet{Lisker2006b}, but which we found to have clearly smaller $g-r$ gradients than $0.10$ mag/$R_{\rm eff}$. {\bf C:} Four normal early-type dwarfs for comparison.}}
\label{fig:colour-profiles-dEbcs}
\end{figure*}

{We find slightly} larger $g-r$ gradients for larger $i-H$ gradients and, while most $i-H$ gradients are negative (median: $-0.116$ mag/$R_{\rm eff}$), $g-r$ gradients are centred around zero with the median at $-0.006$ mag/$R_{\rm eff}$. Additionally, we find a tail of galaxies having clearly larger $g-r$ gradient values. We thus define all 12 galaxies\footnote[5]{VCC~0021, VCC~0173, VCC~0209, VCC~0870, VCC~0951, VCC~1488, VCC~1499, VCC~1501, VCC~1512, VCC~1684, VCC~1779, and VCC~1912} with gradient values larger than 1$\sigma$ above the mean ($\left<\delta(g-r)\right>$= 0.014~mag/$R_{\rm eff}$), i.e. with values above 0.10~mag/$R_{\rm eff}$, as quantitative blue-cored early-type dwarf galaxies {(indicated by the black horizontal tick in Fig.~\ref{fig:gradients-gr-iH})}. Their colours and colour gradients are listed in Table~\ref{tab:dEbcs}. All but one of them (VCC~0209) were already classified as blue-cored early-type dwarfs by the qualitative investigation of \citet{Lisker2006b}, which was based on visual inspection of $g-i$ colour maps. \citet{Lisker2006b} categorized four more {early-type dwarfs} from our sample as {blue-cored} (VCC~0170, VCC~0218, VCC~0308, and VCC~0781), which we find to have clearly smaller $g-r$ gradients. It needs to be noted, however, that VCC~0781 is found to have dust in its centre \citep{diSeregoAlighieri2013}, which might affect the inner colours. For all 12 quantitative {blue-cored early-type dwarf galaxies}, and for the four additional objects in the \citeauthor{Lisker2006b} sample, we show the radial colour profile in Fig.~\ref{fig:colour-profiles-dEbcs}.

For further investigation we divide our sample into four subsamples based on their $g-r$ gradients, such that each subsample includes 30 galaxies, one quarter of the whole sample. {The ranges of these subsamples are shown by the colour-coding in Fig.~\ref{fig:gradients-gr-iH}.} The blue subsample includes the galaxies with the largest, or strongest positive, $g-r$ gradients (bluer centre, redder outskirts), which is the highest quartile, whereas the red subsample includes the galaxies with the smallest, or strongest negative, $g-r$ gradients, which is the lowest quartile.

\begin{figure}
\includegraphics[width=76mm]{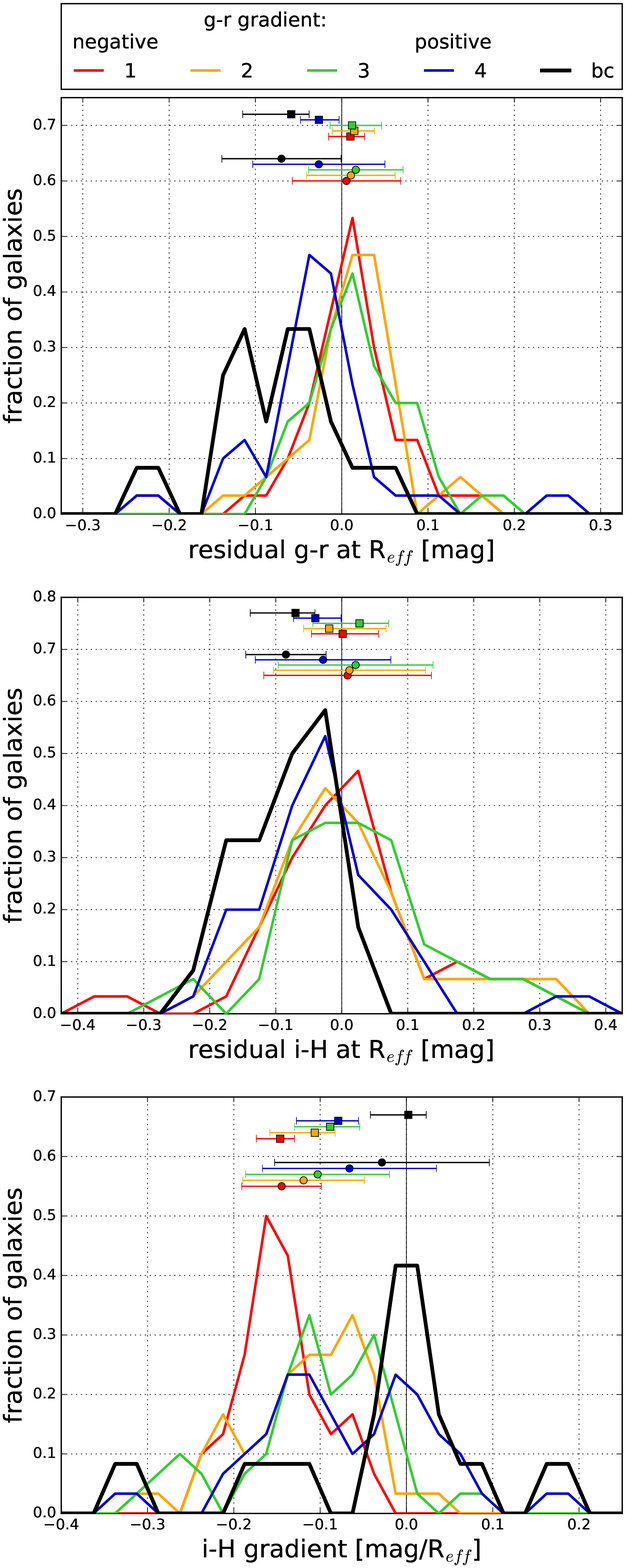}
\caption{{Distribution of $(g-r)_{R_{\rm eff}}$ residual colour (top), $(i-H)_{R_{\rm eff}}$ residual colour (middle), and $i-H$ gradient (bottom), shown separately for each quartile in $g-r$ gradient. These quartiles are 1 = $[-0.100, -0.032]$, 2 = $[-0.031, -0.007]$, 3 = $[-0.004, 0.025]$ and 4 = $[0.028, 0.384]$ in mag/$R_{\rm eff}$. In addition to the four quartiles, the blue-cored early-type dwarfs that we define in Sect.~\ref{subsec:col-colgrad} are shown as black curve. They are also included in the highest quartile (blue curve), i.e. they enter the plot twice. Each curve is normalized to an area of 1 and is drawn by connecting points in steps of 0.05 mag along the x-axis (0.025 mag for the top panel). Each point is the normalized number of galaxies within $\pm0.05$ mag ($\pm0.025$ mag for the top panel). The coloured circles and horizontal bars indicate the respective mean and standard deviation of the distributions, whereas the coloured squares give median and the first and third quartile of the distributions. The {\it uncertainty of the mean value} is smaller by a factor of $\sqrt{30}$ (and by a factor of $\sqrt{12}$ for the blue-cored galaxies) than the displayed standard deviations. }}
\label{fig:delta-gr_3-panels}
\end{figure}

{With} quantitative colour gradients of faint early-type galaxies {at hand, we can address the question of whether gradients are correlated with colours measured at the effective radius and/or the galaxy centre}. In the histogram in Fig.~\ref{fig:delta-gr_3-panels} (top panel) the distribution of the {($g-r$)$_{R_{\rm eff}}$ residual colour} is shown for each of the four subsamples. In addition to the four above-mentioned subsamples, an additional one {comprises only the quantitative blue-cored galaxies (denoted ``bc'' and shown in black in the figure)}. {We find that the blue-cored galaxies tend to have bluer $g-r$ residual colours at the effective radius and therefore even bluer $g-r$ residual colours in the centre than to the rest of the faint early-type galaxies.}

A similar relation is seen in Fig.~\ref{fig:delta-gr_3-panels}, middle panel, where the distribution of the {($i-H$)$_{R_{\rm eff}}$ residual colour} is displayed for the different $g-r$ gradient subsamples. Galaxies with {quantitative blue cores} tend to have bluer $i-H$ resdiual colours at the effective radius. {Owing to} the slight correlation between $g-r$ and $i-H$ gradients {(Fig.~\ref{fig:delta-gr_3-panels}, lower panel)}, {the blue-cored galaxies} also have systematically bluer $i-H$ residual colours at the centre {with respect to the other galaxies of our sample}.

\subsection{{From colours to stellar populations}}
\label{subsec:age-metallicity}

Interpreting our results leads us to the issue of separating age and metallicity effects. It is well known that young stellar populations include hot, massive, short-lived stars that {lead to a bluer integrated} colour and, in contrast, that a high metallicity content {leads to a redder} colour. Therefore, we have an age-colour relation and a metallicity-colour relation, which are both relevant for the interpretation of stellar populations. The question is how strongly each relation {contributes to} a given observed colour. {Based on spectral energy distributions of model stellar populations, optical colours such as $g-r$ can be considered age-sensitive, whereas near-infrared (NIR) colours such as $i-H$ can be considered metallicity-sensitive \citep[see e.g.][]{BruzualCharlot2003}.} 
Therefore, a bluer colour in $g-r$ would imply a younger stellar population, while a redder colour in $i-H$ would imply a more metal-enriched stellar population. 

\begin{figure}
\includegraphics[width=87mm]{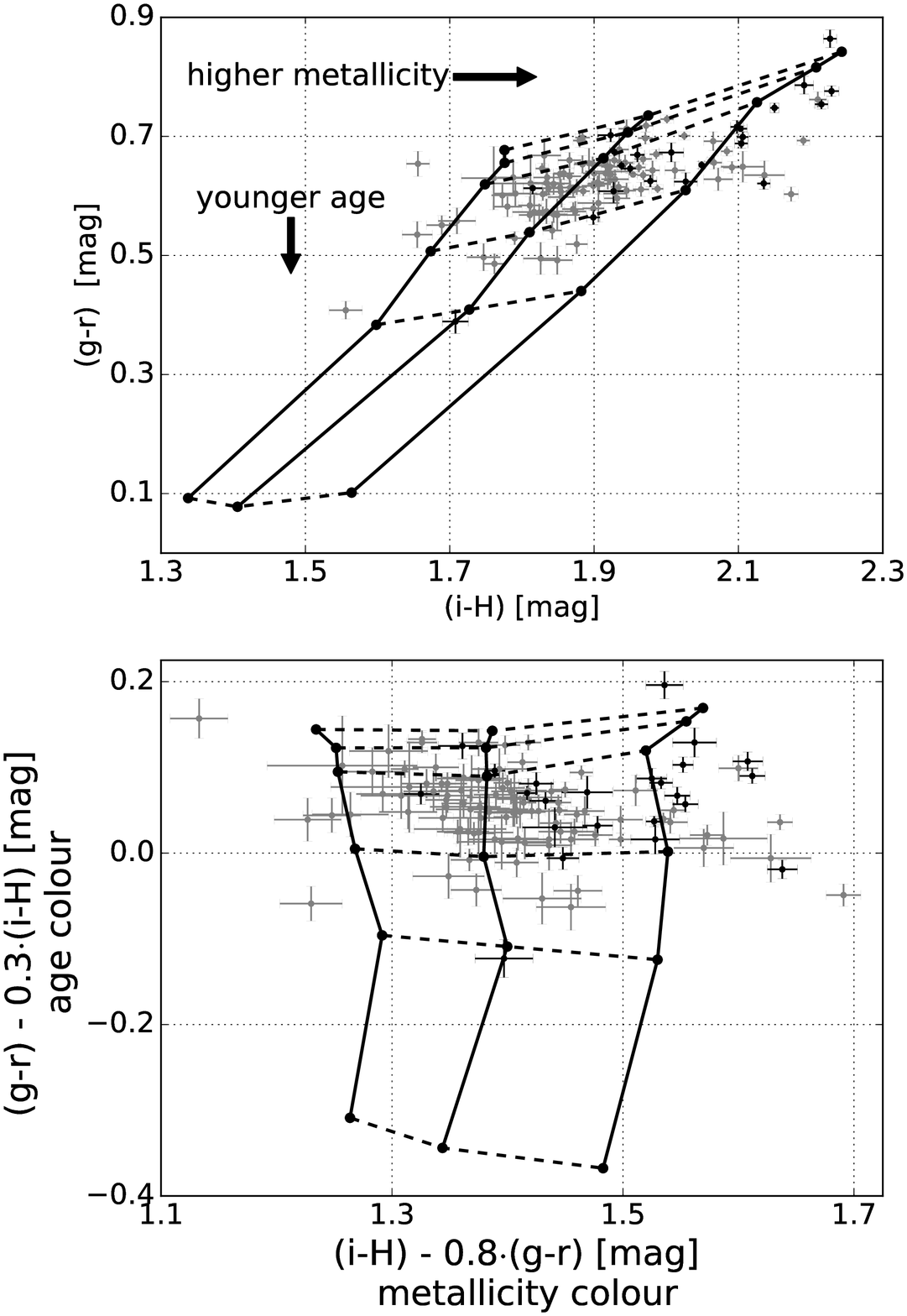}
\caption{{\bf Stellar population model:} Based upon the population synthesis code of \citet{BruzualCharlot2003}, the solid model tracks are curves of constant metallicity (from left to right): Z = 0.008, Z = 0.004, and Z = 0.02, and marked at ages (from bottom to top) of 1, 3, 4.5, 7.5, 10.5, and 13.5~Gyr, where age means the time since the onset of star formation.
{\bf Top panel:} $i-H$ is plotted against $g-r$. 
{\bf Bottom panel:} {\it metallicity colour} $[(i-H)~ - ~0.8~\cdot~(g-r)]$ is plotted against {\it age colour} $[(g-r)~ - ~0.3~\cdot~(i-H)]$. For comparison, {we plot the galaxy colours measured at the effective radius, using grey for galaxies originally classified as dwarfs and black for those originally classified as non-dwarfs.}}
\label{fig:iH-gr_SSP}
\end{figure}

However, looking {in more detail} at the colours of a stellar population model (Fig.~\ref{fig:iH-gr_SSP}) that was generated using the population synthesis code of \citet[][hereafter BC03]{BruzualCharlot2003}\footnote[6]{We use the low-resolution BaSeL 3.1 stellar library, Padova 1994 isochrones \citep{Bertelli1994}, a Chabrier initial mass function \citep{Chabrier2003}, and an exponentially decaying burst of star formation with a decay time $\tau$ = 1~Gyr.}, we find that the $g-r$ colour is mainly age-sensitive and a little bit metallicity-sensitive, whereas the $i-H$ colour is age- and metallicity-sensitive to similar degrees (Fig.~\ref{fig:iH-gr_SSP}, upper panel). Therefore, the slight correlation that we find between $g-r$ and $i-H$ gradients can be explained by age gradients that {contribute equally to} both colours. The trend of $i-H$ gradients being more negative than the corresponding $g-r$ gradients can be explained by metallicity gradients that {contribute more to} $i-H$ than to $g-r$. 

 To generate an age-only and a metallicity-only colour in a simplified way, we subtract the estimated $i-H$ metallicity {contribution} from $g-r$ and the estimated $g-r$ age {contribution} from $i-H$ with appropriate factors (compare the upper and lower panels in Fig.~\ref{fig:iH-gr_SSP}). {{\it We thus define an approximate age-only colour as [$(g-r)~ - ~0.3~\cdot~(i-H)$] and an approximate metallicity-only colour as [$(i-H)~ - ~0.8~\cdot~(g-r)$]};} hereafter we refer to these as {\it age colour} and {\it metallicity colour}.

\begin{figure}
\includegraphics[width=87mm]{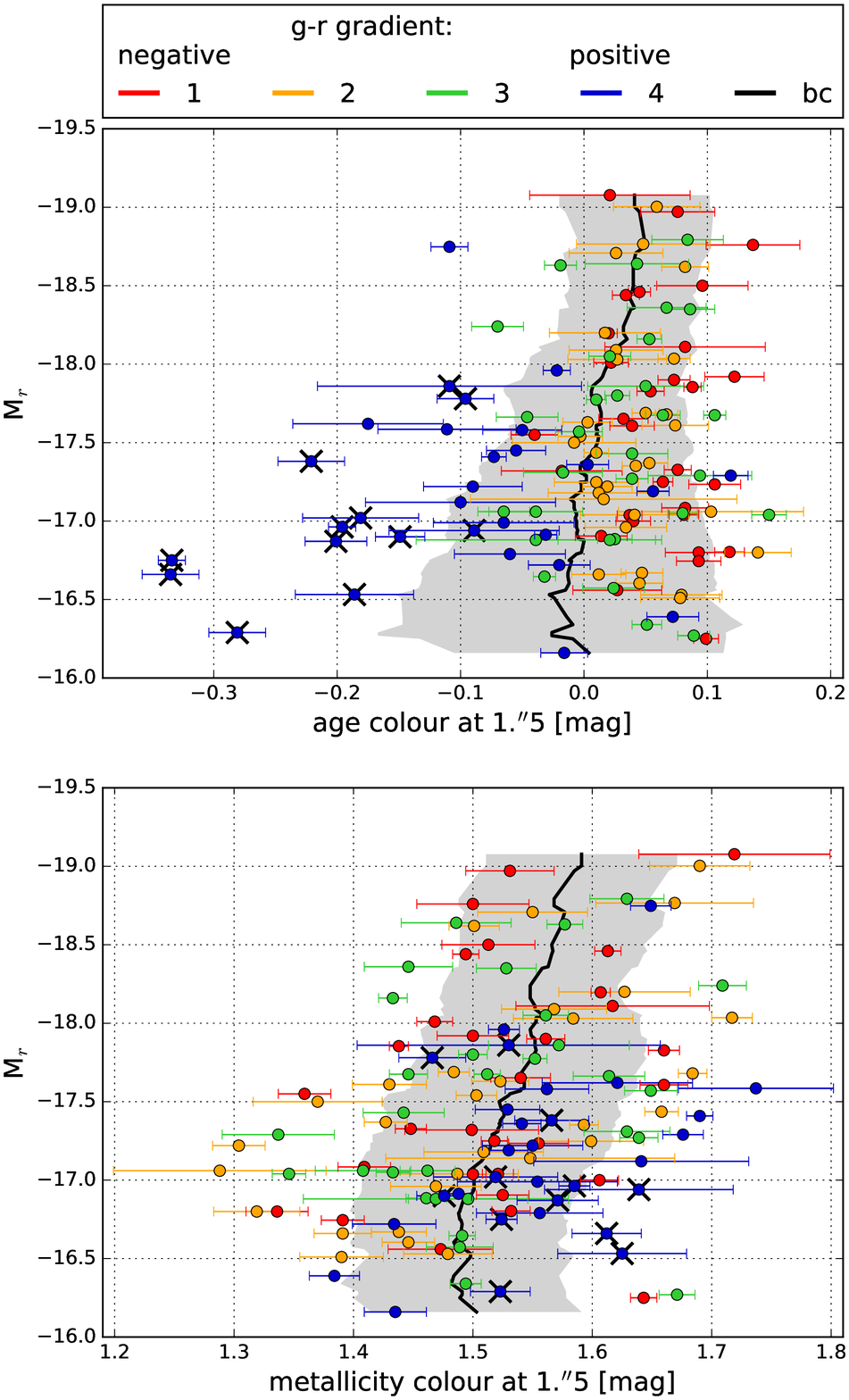}
\caption{{\bf Central colour magnitude relation:} {\it Age colour} measured at 1.$\arcsec$5 (upper panel) and {\it metallicity colour} measured at 1.$\arcsec$5 (lower panel) are plotted against the absolute $r$-band magnitude. {Error bars denote} the uncertainties of the isophotal colours. Galaxies are coloured based on their $g-r$ gradient groups (see Fig.~\ref{fig:delta-gr_3-panels}). The black line describes the {running mean of the colour within a 1~mag interval and the grey shaded area displays the corresponding standard deviation of the galaxies.}}
\label{fig:color-magnitude}
\end{figure}

In Fig.~\ref{fig:color-magnitude} the {central {\it age colour}} (upper panel) and the {central {\it metallicity colour}} (lower panel) are plotted against the absolute $r-$band magnitude. Each galaxy is coloured based on its $g-r$ gradient group ({same colouring as in} Fig.~\ref{fig:delta-gr_3-panels}). Galaxies with stronger positive $g-r$ gradients, especially {blue-cored early-type dwarf galaxies}, tend to have bluer {\it age colours} in the central region, but otherwise we find almost no correlation between {\it age colour} and magnitude. The broad correlation of {\it metallicity colour} and magnitude reflects the well-known relation between stellar mass and metallicity \citep[e.g.][]{BarazzaBinggeli2002}. To take into account these colour-magnitude relations, we computed residual {\it age} and {\it metallicity colours} ({following the approach in Sect.~\ref{sec:analysis}): the difference between a galaxy's colour} and the mean colour (black line in Fig.~\ref{fig:color-magnitude}) of all sample galaxies within a $\pm0.5$ mag interval {around the galaxy is the residual {\it age/metallicity colour} that we use in the following}. 

\begin{figure}
\includegraphics[width=87mm]{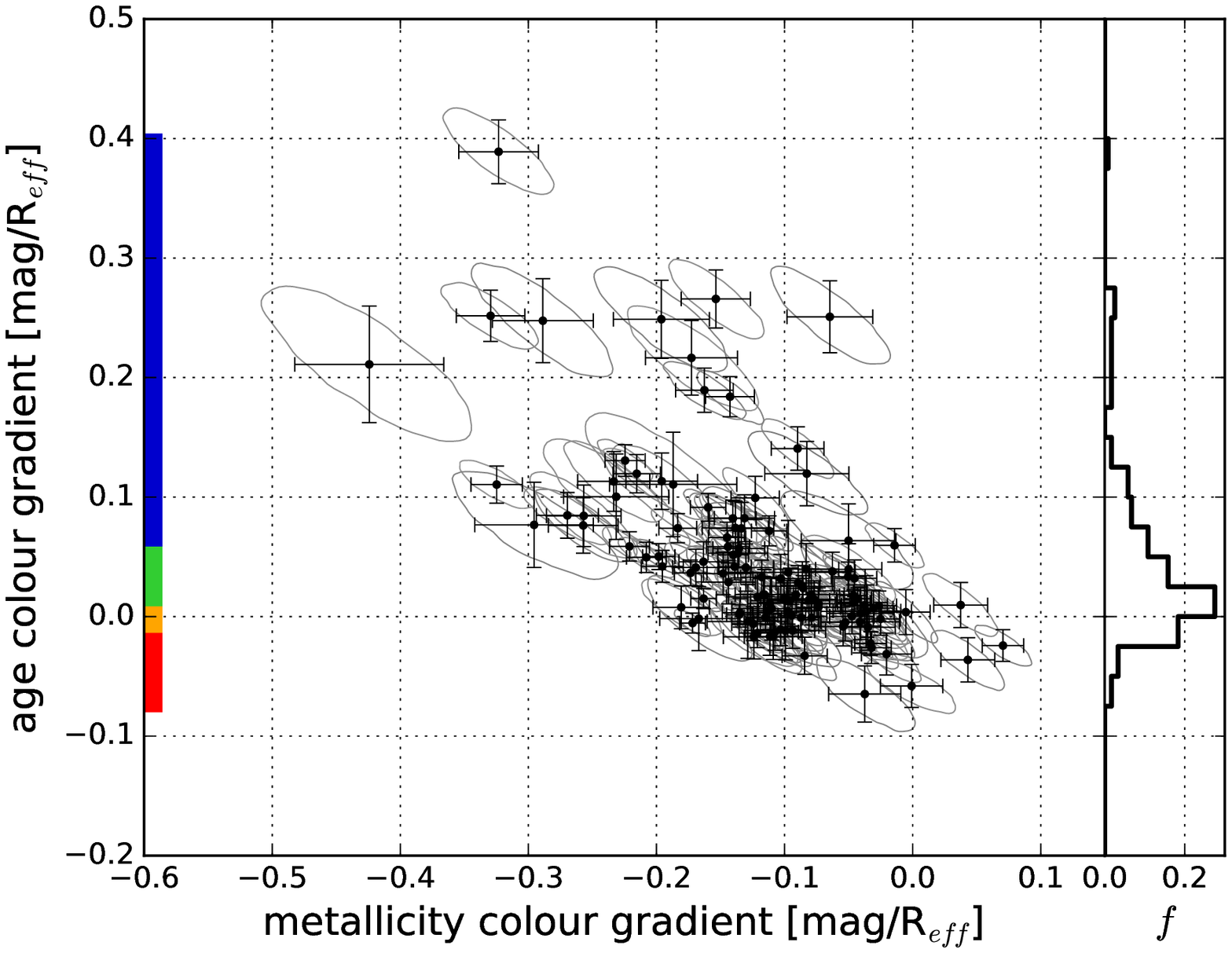}
\caption{{{\it Metallicity colour} gradients of our galaxies plotted against their {\it age colour} gradients (see text for definitions).} {The error bars denote the standard errors (see Sect.~\ref{sec:analysis}).} The correlation of the errors is displayed by a two-dimensional $1 \sigma$ Gaussian contour. The colour-coding on the y-axis defines the ranges of the four subsamples that all include the same number of galaxies (30). The fraction $f$ of galaxies in the {\it age colour} gradient distribution is presented in the right histogram.}
\label{fig:gradients-age-met}
\end{figure}

{Analogously,} we can generate a simplified {\it age colour} gradient of [$\delta(g-r)~ - ~0.3~\cdot~\delta(i-H)$] and a simplified {\it metallicity colour} gradient of [$\delta(i-H)~ - ~0.8~\cdot~\delta(g-r)$]. The resulting values are displayed in Fig.~\ref{fig:gradients-age-met}. {We find that they are the same, within the errors, as when converting each galaxy's $g-r$ and $i-H$ colour profiles into {\it age} and {\it metallicity colour} profiles, and then fitting their gradients.}  

{For the newly defined age/metallicity colours and colour gradients, we have to consider that we introduced correlations by construction and have therefore correlated errors. In Fig.~\ref{fig:gradients-age-met} the correlated errors of the calculated {\it age} and {\it metallicity colour} gradients are displayed as a $1 \sigma$ Gaussian contour.}

\subsection{Stellar populations}
\label{subsec:stellarpop}

\begin{figure}
\includegraphics[width=78mm]{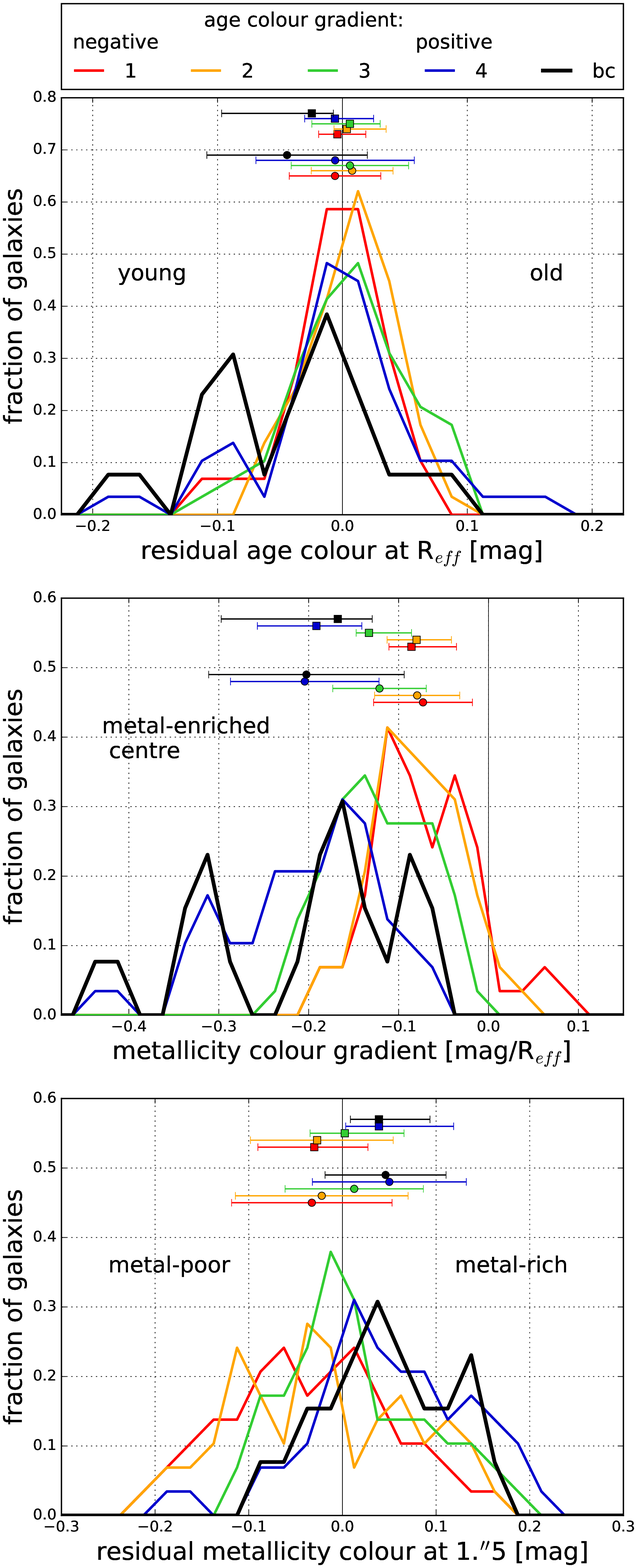}
\caption{{Distribution of residual {\it age colour} (top), {\it metallicity colour} gradient (middle), and residual {\it metallicity colour}  (bottom), shown separately for each quartile in {\it age colour} gradient. These quartiles are 1 = $[-0.065, 0.001]$, 2 = $[0.002, 0.019]$, 3 = $[0.024, 0.072]$ and 4 = $[0.074, 0.389]$ in mag/$R_{\rm eff}$. In addition to the four quartiles, the blue-cored early-type dwarfs that we define in Sect.~\ref{subsec:col-colgrad} are shown as a black curve. They are also included in the highest quartile (blue curve), i.e. they enter the plot twice. Each curve is normalized to an area of 1 and is drawn by connecting points in steps of 0.05 mag along the x-axis. Each point is the normalized number of galaxies within $\pm0.05$ mag. The coloured circles and horizontal bars indicate the respective mean and standard deviation of the distributions, whereas the coloured squares give median and the first and third quartile of the distributions. The {\it uncertainty of the mean value} is smaller by a factor of $\sqrt{30}$ (and by a factor of $\sqrt{12}$ for the blue-cored galaxies) than the displayed standard deviations. }}
\label{fig:diff-grad-age-hist}
\end{figure}

In Fig.~\ref{fig:diff-grad-age-hist} the sample is divided into four equal subsamples based on their {\it age colour} gradients. The ranges of these subsamples are shown by the colour-coding in Fig.~\ref{fig:gradients-age-met}. The red subsample includes galaxies with negative {\it age colour} gradients (the lowest quartile) and the blue subsample includes the galaxies with the strongest positive {\it age colour} gradients (the highest quartile). The black subsample contains all quantitative blue-cored early-type dwarf galaxies, as before, and all of them are {also} part of the blue subsample. For each subsample the distribution of the residual {\it age colour} measured at the effective radius is plotted in the upper panel of Fig.~\ref{fig:diff-grad-age-hist}. We find no clear correlation between the {\it age colour} gradient and the residual ({\it age colour})$_{R_{\rm eff}}$, but we find that the galaxies defined as quantitative {blue-cored early-type dwarf galaxies} tend to have bluer {residual ({\it age colours})$_{R_{\rm eff}}$}. This implies that {blue-cored early-type dwarf galaxies} tend to have a younger stellar population overall, and an even younger stellar population in their central region, since {blue-cored early-type dwarf galaxies} possess strong {\it age colour} gradients. 

In the middle panel of Fig.~\ref{fig:diff-grad-age-hist}, where the distribution of the {\it metallicity colour} gradient is plotted for each subsample, we see that nearly all galaxies have negative {\it metallicity colour} gradients. {Galaxies} with a strong {\it age colour} gradient also tend to have a strong {\it metallicity colour} gradient. Furthermore, we find that the galaxies with the strongest {\it age colour} gradients possess the reddest {central residual {\it metallicity colours}} (Fig.~\ref{fig:diff-grad-age-hist}, lower panel). These results imply metal-enriched central regions for almost all early-type dwarf galaxies and, at first glance, that galaxies with strong {\it age colour} gradients also have the strongest metallicity gradients and the most metal-enriched stellar populations in their centre. 

 However, we have to consider that colours and colour gradients are luminosity weighted. {Therefore, we need to consider} a more realistic stellar composition that consists of a younger and an older population.

\begin{figure}[]
\includegraphics[width=87mm]{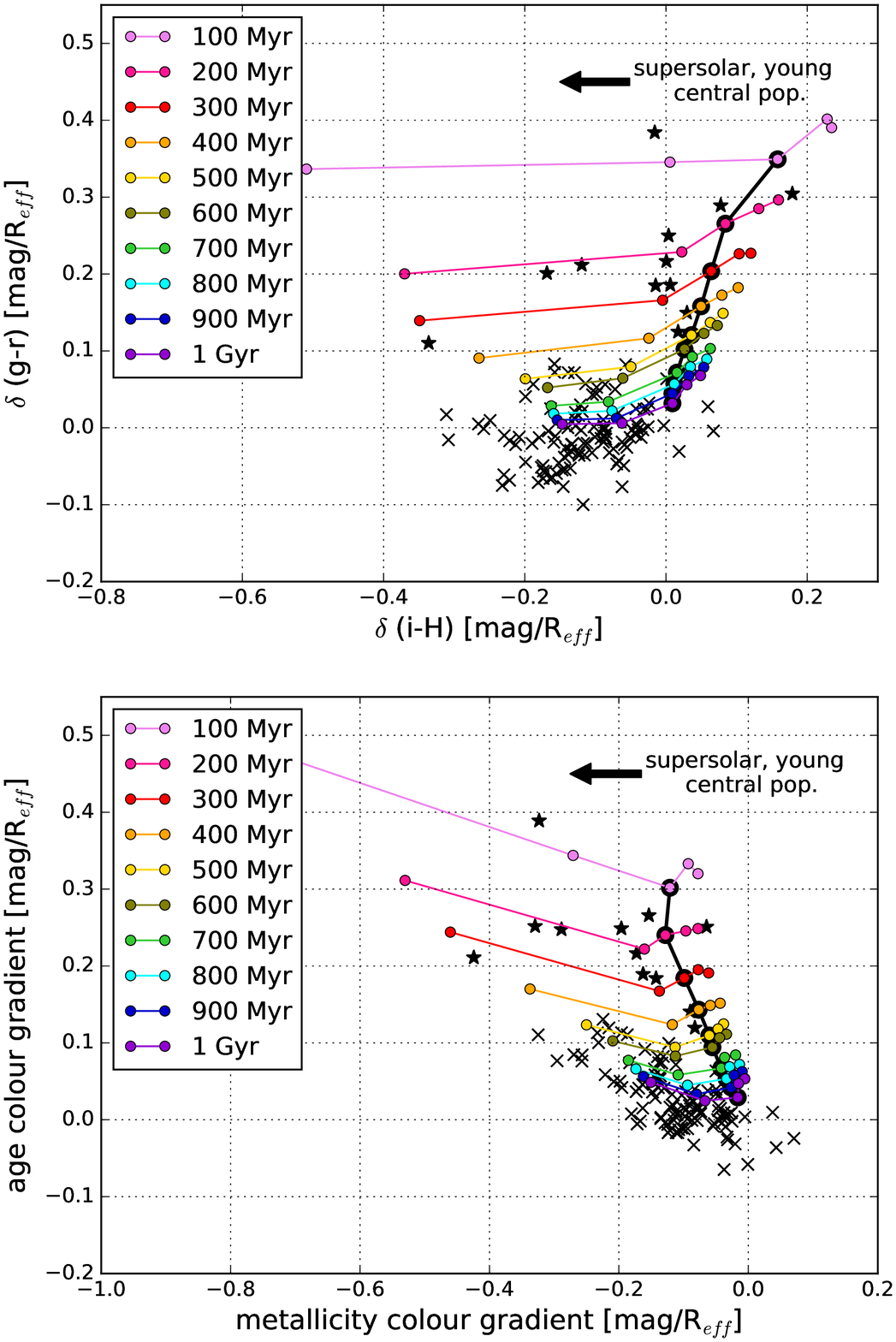}
\caption{{Composite stellar population model,} {based upon the} population synthesis code of \citet{BruzualCharlot2003}. {Each model track has a single value for the age gradient} (old pop.: 7~Gyr, Z = 0.004; composite pop.: 95\% old and 5\% young; age of young pop. in the label) and is marked at different metallicity gradients (metallicity of the young pop. from left to right: Z = 0.1, Z = 0.05, Z = 0.02, Z = 0.008, and Z = 0.004). The black solid line {denotes} solar metallicity for the young population. {Blue-cored early-type dwarf galaxies are displayed as asterisks.}}
\label{fig:grad-ssp}
\end{figure}

To compare our colour gradient values with synthetic colour gradients modelled from composite stellar populations we used the BC03 library. Following the simplified approach of \citet{Lisker2006b} of using single-age populations, we assume an old population of 7~Gyr and Z = 0.004 for the whole galaxy (estimated from Fig.~\ref{fig:iH-gr_SSP}) and an additional young population that makes up 5\% of the local {present-day} stellar mass in the centre (discussed in section \ref{subsec:stellar-content}). We varied age and metallicity of the young population {and} calculated the colour difference {between} the old population ({which represents the situation at $R_{\rm eff}$}) and the composite population ({which represents the situation at the galaxy centre}). {We use these calculated differences as model gradients and compare them to our fitted gradients (Fig.~\ref{fig:grad-ssp})}. {Each model track in the figure has a single value for the age gradient, with} the age of the young population given in the legend. The tracks are marked at different metallicity gradients, {with} the metallicity of the young population from left to right: Z = 0.1, Z = 0.05, Z = 0.02, Z = 0.008, and Z = 0.004. Therefore, the rightmost point {of a model track} represents an age-only gradient and the leftmost point represents the largest metallicity gradient in addition to the age gradient. 

{We find that a stellar population younger than 500~Myr is needed to match the measured colour gradients of the blue-cored early-type dwarf galaxies in addition to assuming a mean age of 7~Gyr and a mean metallicity of ~Z = 0.004.} When we increase the {present-day mass} ratio of the young population up to 9\%\footnote[7]{\citet{Lisker2006b} assumed that the old population makes up at least 90.9\% and the young population less than 0.3\% of the total mass, {except for} VCC~1499 for which they assume that the old population makes up 75\% and the young population 0.8\%.} we can reproduce the {blue-cored early-type dwarfs'} colour gradients with a young population {of up to} 700~Myr. Furthermore, when we increase the mean age to 9~Gyr, the young population has to be {younger} than 600 Myr, and when we decrease the mean age to 5~Gyr, we need a young population younger than 400~Myr to {cover the locus of blue-cored early-type dwarfs in Fig.~\ref{fig:grad-ssp}}. 
Additionally we find that most {blue-cored early-type dwarf galaxies}, as well as most normal {early-type dwarfs}, have a central stellar population with supersolar metallicity. This, however, does not imply a supersolar {\it mean} metallicity since the central {metal enriched (young)} stellar population makes up only a small fraction of the total mass. 

As mentioned previously, we have to consider that colour gradients are luminosity weighted. The younger the young population, the larger its contribution to the {galaxy's central colour} and therefore, the stronger the colour gradients.
This explains why, in Fig.~\ref{fig:grad-ssp}, the vertical steps between the model tracks {become larger} with decreasing age of the young population, even though the step in {the age gradient itself is always exactly 100~Myr}. It also explains why the tracks become longer horizontally, even though the difference in the young population's metallicity between the left and right ends of the tracks is always the same. 

 In summary, we can say that almost all early-type dwarf galaxies have non-zero metallicity gradients and that the metallicity gradients of {blue-cored early-type dwarfs} are consistent with those of normal {early-type dwarfs}. 

%-----------------------------------------------------------------------------------
%-----------------------------------------------------------------------------------
\section{Discussion}
\label{sec:discuss}

\subsection{Stellar content}
\label{subsec:stellar-content}

Our colour analysis indicates that the stellar content of blue-cored early-type dwarf galaxies can be explained by a young stellar population that makes up a small fraction of the mass in the central region. Therefore, {blue-cored early-type dwarf galaxies} are similar to normal {early-type dwarfs}, but have an additional young stellar population, or star forming region, in the centre. {This is confirmed by the presence of prominent Balmer absorption in most cases and H$\alpha$-emission in some cases, as noted by \citet{Lisker2006b} from SDSS spectra of the inner 1.$\arcsec$5 (in radius). These authors modelled\footnote[8]{They used the BC03 library with high spectral resolution, which was produced using the \citet{BruzualCharlot2003} Galaxy Isochrone Spectral Synthesis Evolution Library (GISSEL) code, implementing the Padova 1995 isochrones \citep{Fagotto1994,Girardi1996} combined with the STELIB \citep{LeBorgne2003} stellar library.} the spectra of blue-cored early-type dwarfs} with a composite stellar population made up of
three simple stellar populationss: an old (5~Gyr), an intermediate-age (10~Myr to 1~Gyr), and a young (<10~Myr) stellar population. Furthermore they assumed the same metallicity (Z = 0.004) for all populations.

However, their best-fitting composite populations would have systematically bluer $i-H$ colours and systematically more positive $i-H$ gradients than our measured values. The $g-r$ colour and colour gradients would be in the same overall range, but with a tendency toward slightly bluer colours and more positive gradients. These apparent discrepancies in $i-H$ can be explained by the lack of a metallicity difference between the old and young populations of \citet{Lisker2006b}. {When} assuming a higher metallicity for the young and maybe the intermediate-age population we get redder $i-H$ colours and weaker $i-H$ gradients. Additionally, we have to consider that \citeauthor{Lisker2006b} investigated only the innermost region of 1.$\arcsec$5 with their spectral analysis, while we do not examine this central part in our colour analysis. Since we found that {blue-cored early-type dwarfs} possess strong age gradients, it is possible that \citeauthor{Lisker2006b} observed a very young stellar population in the central part that our analysis is not sensitive to.

 Additionally, \citet{Lisker2006b} mentioned that varying the ages of the different populations has a similar effect on the composite spectrum to changing the ratio of the mass fractions. We find the same result for the colours and colour gradients. When we assume an old population of 7~Gyr that makes up 95\% of the total mass and a young population of 200~Myr, the colours and gradients are similar to when we assume instead an old population of 7 Gyr that only makes up 91\% of the total mass and a young population of 300~Myr, or an old population of 9~Gyr that makes up 95\% of the total mass and a young population of 300~Myr.

Therefore, it is not possible to determine the exact stellar content of blue-cored and normal early-type dwarf galaxies, especially as we only perform a colour analysis and do not have spectra of different regions within a galaxy. {Nevertheless,} we can conclude that normal early-type dwarf galaxies have a mean age of about 5-9~Gyr and a subsolar mean metallicity {within} the framework of \citeauthor{BruzualCharlot2003}'s stellar population models, as described in section \ref{subsec:age-metallicity}. This is consistent with the results of e.g. \citet{Geha2003} and \citet{vanZee2004a} who found in their spectral analysis a mean age of 5-7~Gyr and clearly subsolar metallicities, and the results of \citet{Michielsen2008} and \citet{Toloba2014b} who found a mean age of about 6~Gyr and a mean metallicity of {$log(Z/Z_{\odot}) \approx -0.6$. \citet{Roediger2011}} found a somewhat older mean age (9-10~Gyr) of {dwarf ellipticals} from a different data set of optical-NIR photometry. {When we subdivide our sample of faint early-type galaxies into galaxies originally classified as dwarfs and non-dwarfs, we find a similar age range for both, but a trend for non-dwarfs to be a bit more metal-rich than the dwarfs (see Fig.~\ref{fig:iH-gr_SSP}). This is consistent with the findings of \citet{Guerou2015}.} 

 We can infer that blue-cored early-type dwarf galaxies have {mean ages and metallicities within the range of normal faint early-type galaxies, and are not separated from them. They do, however,} belong to those {galaxies} with an overall younger stellar age. They feature an additional stellar population in the central region that has supersolar metallicity and is younger than 500~Myr, which implies that a blue-cored {early-type dwarf galaxy} can become a normal {early-type dwarf} within 0.5 Gyr, if no further central star formation occurs. In the literature it is often said that early-type dwarf galaxies have solar or subsolar metallicities \citep[e.g.][]{vanZee2004b}. However, \citet{Gu2006} found a high-metallicity H\RM{2} region in the blue-cored dwarf elliptical galaxy IC 225 and \citet{Peeples2008} found some low-mass galaxies with high central gas metallicity and blue cores. \citet{Peeples2008} predicted that if a galaxy has a relatively low gas fraction, a small amount of pollution is able to enrich the gas. They suggested that these galaxies are transition types from gas-rich dwarf irregulars to gas-deficient dwarf ellipticals, even though it remains unclear whether a significant fraction of stars would still be able to form from the enriched gas. However, supersolar metallicities in {faint early-type galaxies} are already observed and are not inconsistent with a mean subsolar metallicity, since the high-metallicity portion makes up only a small fraction in mass.
 
In addition, we found that nearly all {early-type dwarfs} possess negative metallicity gradients, which has also been found by \citet{Sybilska2017}, and across all galaxy types in the Virgo Cluster \citep[e.g.][]{Roediger2011}. The {lower} panel of Fig.~\ref{fig:iH-gr_SSP} indicates that a metallicity difference of about 0.7~dex (left to right model track) implies a colour difference of more than 0.2~mag in {\it metallicity colour}, while a metallicity difference of 0.3~dex (left to middle model track) implies a colour difference in {\it metallicity colour} of about 0.1~mag. Since we find that most {early-type dwarf galaxies} possess a {\it metallicity colour} gradient of about $-0.1$~mag/$R_{\rm eff}$ and almost all {blue-cored early-type dwarf galaxies} a {\it metallicity colour} gradient $\lesssim -0.1$~mag/$R_{\rm eff}$, our results are consistent with the average value of $-0.5$~dex/$R_{\rm eff}$ for dwarf elliptical galaxies in the Fornax cluster and nearby groups of galaxies found by \citet{Koleva2009}. The metallicity gradient for the Local Group dwarf elliptical galaxy NGC 205 of $-0.6$~dex/$R_{\rm eff}$ found by \citet{KolevaPhD2009} and the values for Virgo cluster early-type dwarf galaxies found by {\citet[][his table 3]{Chilingarian2009}} are also consistent with our results. 

In our whole analysis we neglected dust extinction, {which in principle could cause redder colours. If dust had a significant effect on our galaxies' colours, we should find the flattest galaxies to be redder, which is not the case; in fact, we find that the roundest galaxies tend to be redder in $i-H$.} Additionally, \citet{Conselice2003}, \citet{diSeregoAlighieri2007}, and \citet{DeLooze2010} showed that early-type dwarf galaxies include little to no dust. {\citeauthor{DeLooze2010}} argued that the removal of interstellar dust from dwarf galaxies has to be as efficient as the removal of interstellar gas.  Nonetheless, VCC~0209, the galaxy with the strongest {\it metallicity colour} gradient ($-0.42$ mag/$R_{\rm eff}$) has been detected with WISE \citep{Boselli2014} and Herschel \citep[][in all five bands]{Auld2013} and therefore contains a significant amount of dust.

{Another factor which has to be considered in blue-cored galaxies is the projection effect. Ignoring any gas and dust content we can make a simple estimate for the integrated amount of light that we receive along a given line of sight. We adopt an exponential profile and consider a flattened dwarf galaxy with an effective radius of 1.2~kpc along the major axis (median size of our blue-cored dwarf sample) and an axis ratio of 0.45 \citep[which is approximately the intrinsic axis ratio of blue cored early-type dwarfs that was inferred by][]{Lisker2007}. We assume that the galaxy includes a young population that is confined to an inner region with a semi-major axis of 0.4~kpc and an axis ratio of 0.85, i.e.\ rounder than the main body of the galaxy. We now compare the situation when observing this galaxy along its minor and its major axes, with a line of sight through the galaxy centre. The light fraction $f$ that we would receive from the inner region along the minor axis ($f = 0.65$) is 1.5 times larger than the fraction we would get along the major axis ($f= 0.43$). This factor increases to 1.7 if we halve the size of the inner region; it decreases to 1.4 if we flatten the inner region to an axis ratio of 0.7. If we now assume that the young population makes up 7.7\% of the present-day stellar mass of the inner region, it would contribute 5.0\% to the light along the minor axis and 3.3\% along the major axis. A $g-r$ gradient that we would measure to be 0.10 mag/R$_{\rm eff}$ when observed along the minor axis would therefore be measured to be only 0.075 mag/R$_{\rm eff}$ when observed along the major axis. In principle, this could lead to missing detections of quantitative blue-cored early-type dwarf galaxies, which we defined as having more than 0.10 mag/R$_{\rm eff}$. However, in our sample there are only two galaxies, VCC~0165 and VCC~1049, with gradients in this range.}

\subsection{Formation scenarios}

It has often been suggested that {early-type dwarf galaxies} are remnants of infalling late-type galaxies that are modified by the strong forces acting within a cluster environment \citep[e.g][]{vanZee2004b,Boselli2008,DeRijcke2010,Toloba2014a} and that {blue-cored early-type dwarf galaxies} may be transition types between late-type galaxies and {early-type dwarfs} \citep[e.g.][]{Lisker2006b}. A strong age gradient with a younger, metal-enriched central region could in principle have been generated by quenching star formation only in the outskirts and/or by increasing the star formation rate in the centre. Additionally, if part of the {early-type dwarf galaxies} had joined the Virgo cluster more recently, we would also expect a younger {\it mean} age. This is exactly what we find for the {blue-cored early-type dwarf galaxies}. 

 The two most often suggested scenarios are ram pressure stripping \citep{GunnGott1972} and galaxy harassment \citep{Moore1996,Moore1998}; the latter can be split into two components, tidal shocking with a short timescale as in high-speed, massive encounters and tidal heating with a long timescale as for galaxies that pass the central cluster's potential well closely \citep{Smith2010}. \citet{Moore1998} showed in their simulations that galaxy harassment causes the gas to funnel to the centre and form a gas density excess. However, it remained unclear whether this would lead to blue cores, as the authors remarked that feedback could halt the collapse of the gas. Moreover, \citet{Smith2010} found in a Monte Carlo simulation of galaxy harassment by a Virgo-like harasser population that strong tidal encounters are very rare, involving less than 15\% of the infalling dwarfs. Furthermore, \citet{Smith2015} {pointed out} that although their {model galaxies} spent many Gyrs in the cluster environment, most dwarf galaxies showed no stripping or change in radial distribution of stars. Hence, it has not been ruled out that galaxies are influenced by harassment, but if galaxy harassment is able to trigger star formation it would only be efficient for a small fraction of the dwarf galaxies in a Virgo-like cluster, and those would be in a range of different evolutionary states at present. It should be noted, however, that the slower tidal interactions typical of group environments were found to be able to induce gas infall and cause central starbursts in simulations of low-mass disk galaxies \citep{YozinBekki2015}. {This may be of relevance if the Virgo cluster {blue-cored early-type dwarf galaxies} -- possibly an infalling population \citep{Lisker2006b} -- had resided in groups until recently and were pre-processed there \citep[e.g.][]{DeLucia2012}.}

There is more agreement with the ram pressure stripping scenario, where the interstellar medium (ISM) of a late-type galaxy is removed by the hot intracluster medium (ICM) as it flows past. For example, \citet{Conselice2001} argued that the Virgo cluster dwarfs are an infall population and \citet{Boselli2008} {demonstrated that} ram-pressure stripping {can play} a prominent role {for} the star formation histories of dwarf galaxies if gas is removed quickly. Furthermore, \citet{DeRijcke2010} showed that some observational quantities in the Fornax cluster could be explained by assuming late-type dwarf galaxies falling into the cluster and evolving into early-type dwarf galaxies by ram pressure stripping. \citet{MoriBurkert2000} found that for galaxies with masses {below} $10^{9}M_{\odot}$, which is roughly the mass of most {blue-cored early-type dwarf galaxies, the ram pressure of the ICM in a typical cluster environment} exceeds the gravitational force of the galaxy. {Thus their gas gets removed entirely, but for} more massive galaxies, the gas can survive in the central region of the gravitational potential well. Therefore, the gas in the outer regions of the infalling galaxy can be stripped by the ICM, while the gas in the inner part remains. {It can} be shown that the Virgo intracluster medium \citep{Vollmer2009} would not be able to remove gas from the centres of most quantitative blue-cored early-type dwarf galaxies.\footnote[9]{At a clustercentric distance of 0.5 Mpc the density of the Virgo intracluster medium as modelled by \citet{Vollmer2009} provides a ram pressure of $10^{-11.13}$ N/m$^2$ when assuming a relative velocity of 1000 km/s, and a ram pressure of $10^{-11.32}$ N/m$^2$ at a clustercentric distance of 1 Mpc.
We use $ugriz$ photometry to approximate the stellar surface mass density (based on \citet{Hansson2012} at one exponential scale length) for all quantitative blue-cored early-type dwarf galaxies in the Virgo cluster. When assuming a gas surface density of one-tenth of the stellar surface density we obtain estimates for the restoring force per unit area between $10^{-11.94}$ and $10^{-10.76}$ N/m$^2$ {and between $10^{-11.08}$ and $10^{-9.89}$ N/m$^2$ when extrapolating these values to the galaxies' centres, using an exponential profile.}} Thus, the galaxy continues to have active star formation in the central region -- at a rate that may even be increased due to ram pressure \citep{Kronberger2008} -- while the outer parts are reduced to a passively evolving stellar population. Such a development clearly implies an age gradient and, if the central region retains a large fraction of its enriched material, also a metal-enriched centre. Since this is exactly what we find, our colour analysis corroborates the idea of ram pressure stripping.

It remains to be mentioned that it is also possible that some {early-type dwarf galaxies} do not originate from infalling galaxies and that not all {early-type} dwarfs have necessarily been {blue-cored early-type dwarfs} at some point in the past. 
To get a better insight into their evolution, it is important to have larger samples of spectroscopic [$\alpha$/Fe] abundances for low-mass early-type galaxies in order to understand how the (early) environment governed the star formation timescales of low-mass galaxies \citep{Michielsen2008,Liu2016}, and hence their stellar population build-up. From simulations, \citet{Schroyen2011} showed that the development of metallicity gradients is anticorrelated with angular momentum \citep[see also][]{Leaman2013}. Accurate measurements of stellar age and metallicity gradients \citep{Sybilska2017} in conjunction with the angular momentum distribution from integral-field spectroscopic data \citep{Rys2014,Guerou2015} are essential for {early-type dwarf galaxy} samples of significant size over a range of environments and, ideally, stellar mass.

%-----------------------------------------------------------------------------------
%-----------------------------------------------------------------------------------
\begin{acknowledgements}
     TL and RFP acknowledge financial support from the European Union's Horizon 2020 research and innovation programme under the Marie Skłodowska-Curie grant agreement No 721463 to the SUNDIAL ITN network. 
     GvdV, RL, AS, RFP and JF-B acknowledge support from grant AYA2016-77237-C3-1-P from the Spanish Ministry of Economy and Competitiveness (MINECO).
     Paudel S. acknowledges the support from the Samsung Science \& Technology Foundation under Project Number SSTF-BA1501-0.
\end{acknowledgements}

\end{document}